\RequirePackage{lineno}
\documentclass[aps,prl,twocolumn,groupedaddress,showpacs,letterpaper,floatfix,10pt]{revtex4-1}
\usepackage{amsfonts}
\usepackage{amsmath}
\usepackage{amssymb}
\usepackage{graphicx}
\usepackage{natbib}
\usepackage[colorlinks=true]{hyperref}
\usepackage[noend]{algpseudocode}
\usepackage{newfloat}
\DeclareMathAlphabet\mathbfcal{OMS}{cmsy}{b}{n} 
\DeclareFloatingEnvironment[
    fileext=loa,
    listname=List of Algorithms,
    name=ALGORITHM,
    placement=tbhp,
]{algorithm}

\begin{document}
\title{Validity of the Local Approximation in Iron- Pnictides and Chalcogenides}
\author{Patrick S\'emon}
\affiliation{Department of Physics and Astronomy, Rutgers University, Piscataway, NJ 08854, USA}
\author{Kristjan Haule}
\affiliation{Department of Physics and Astronomy, Rutgers University, Piscataway, NJ 08854, USA}
\author{Gabriel Kotliar}
\affiliation{Department of Physics and Astronomy, Rutgers University, Piscataway, NJ 08854, USA}
\affiliation{Condensed Matter Physics and Materials Science Department, Brookhaven National Laboratory, Upton, NY 11973-5000, USA}
\begin{abstract}
We  introduce a  cluster DMFT (Dynamical Mean Field Theory)  approach to study the normal state of the iron pnictides and
chalcogenides. In the regime of moderate mass renormalizations, the self-energy is
very local, justifying the success of single site DMFT for these materials and for other
Hunds metals. We solve the corresponding impurity model with CTQMC (Continuous Time Quantum Monte-Carlo) and find that
the minus sign problem is not severe in regimes of moderate mass renormalization. 
\end{abstract}

\maketitle


The unexpected discovery of superconductivity in the iron pnictide based materials has opened a new era of research in the field of condensed matter physics.\cite{Hosono:2006} Multiple approaches, starting from weak coupling such as the random phase approximation (RPA) and strong coupling approaches using lessons learned from the t-J model, have been proposed, but there is not yet consensus in the community of what constitutes the proper theoretical framework for describing these systems.\cite{ChubukovHirschfeld:2014} It has been proposed that iron pnictides and chalcogenides are important not only because of their high temperature superconductivity, but because their normal state properties represent a new class of strongly correlated systems, the Hunds metals.  They are distinct from doped Mott Hubbard systems, in that correlations effects in their physical properties derive from the Hunds rule coupling J, rather than the Hubbard U. \cite{HauleKotliar:2009, YinHauleKotliar:2011} Many other interesting Hunds metals have been recognized, as for example Ruthenates \cite{Mravlje:2011} and numerous 3d and 4d compounds \cite{Antoine:2013}.

Dynamical Mean Field Theory\cite{Georges:1996}(DMFT) and its cluster extensions\cite{Kotliar:2001,MaierRev:2005} have provided a good  starting point for the description of Mott Hubbard physics. It is now established that it describes many puzzling properties of three dimensional materials such as Vanadium oxides near their finite temperature Mott transition.\cite{Deng:2014} In materials such as cuprates, as  the temperature is lowered,  the description in terms of single site DMFT gradually breaks down.  New  phenomena such as  momentum space differentiation and the opening of a pseudogap takes place,\cite{Huscroft:2000, Lichtenstein:2000,Jarrell:2001a, Jarrell:2001b, Haule:2003, Parcollet:2004, Carter:2004, Civelli:2005,Stanescu:2006,Kyung:2006a, Macridin:2006, Haule:2007b, Zhang:2007,Civelli:2008,Civelli:2009,Liebsch:2009,Sakai:2009,Werner:2009,Gull:2010, Lin:2010,Sordi:2012,Sordi:2013,Gull:2013a,Gull:2013b,Imada:2013} and cluster DFMT is essential. How different cluster sizes and methods captures these effects has been explored intensively.\cite{Jarrell:2001b,Biroli:2002,Aryanpour:2005,Biroli:2005,Maier:2005b,Kyung:2006b,Gull:2010,Sakai:2012}
The iron pnictides and chalcogenides have been extensively studied using LDA+DMFT by several groups.\cite{HauleKotliar:2009, YinHauleKotliar:2011, Yin:2014, Valenti:2015, Aichhorn:2010} It has been argued using the GW method, that the frequency dependence of low order diagrams in perturbation theory in these materials is very local.\cite{Tomczak:2012} However, because of the difficulties posed by the multiorbital nature of these compounds, the accuracy of the local approximation beyond the GW level has not been examined and is the main goal of this paper. 

Building on the work of Ref.~\onlinecite{Ferrero:2009}, we introduce a cluster extension for the treatment of iron pinctides, which is numerically tractable using CTQMC. By comparing single site and cluster DMFT, we  establish that in a broad range of parameters where the mass renormalizations are of the order of 2 to 3, which corresponds to the experimental situation in many iron pnictides and chalcogenides, the local approximation is extraordinarily accurate, justifying the success of a very large body of work.


For simplicity, we use in this work a tight-binding hamiltonian $\mathbf{h}_0(\mathbf{k})$ of $FeAs$ layers with $As$ treated in second order perturbation theory, as presented by M. J. Calder\'on et al.\cite{Calderon:2009} For the hopping amplitudes the values suggested for $LaOFeAs$ are taken and scaled such that the bandwidth is $\approx 4eV$.\footnote{This corresponds to choose $(pd\sigma)^2/|\epsilon_d - \epsilon_p|=0.75eV$ in Ref.~\onlinecite{Calderon:2009}.} However, the main conclusions of this work should not be very sensitive to the parametrization used.
The wave vectors $\mathbf{k}$ label the irreducible representations of a glide-mirror symmetry group instead of the usual translation symmetry group, so that the Brillouin zone contains 1 $Fe$ atom instead of 2 $Fe$ atoms, with hole pockets at the $M$ and $\Gamma$ points and electron pockets at the $X$ and $Y$ points. Notice here that this unfolding, which is exact in two dimensions, is not exact when the $FeAs$ layers are coupled, i.e., a translation operation perpendicular to the layers does not commute with a glide mirror operation along the layers, and the corresponding symmetry group is not abelian. The correlations of the electrons within a $d$-shell are captured by adding a local Coulomb interaction, parametrized by the Hubbard repulsion $U$ and the Hund's rule coupling $J$, see supplementary information Sec.~D for more details.

\begin{figure}[h]
\centering
\includegraphics[width=1.0\columnwidth]{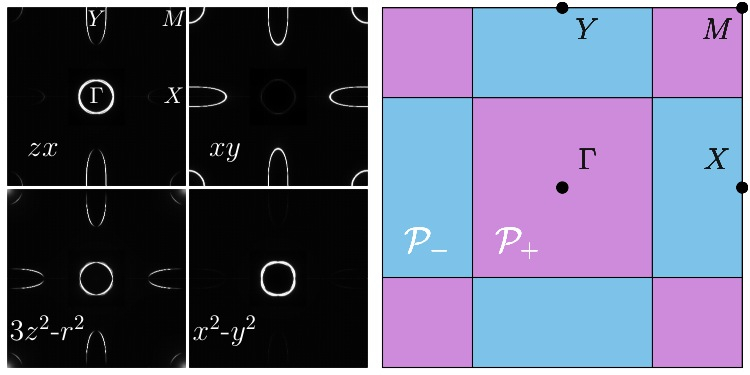}
\caption{Left panel: Orbital character of the Fermi surface in the unfolded Brillouin zone of the tight-binding hamiltonian\cite{Calderon:2009} used in this work. Right panel: Tiling of the Brillouin zone in two patches $\mathcal{P}_+$ and $\mathcal{P}_-$, enclosing the holes pockets at $\Gamma$ and $M$ and the electron pockets at $X$ and $Y$ respectively. This patching is compatible with the lattice symmetries.}
\label{fig:BrillouinPatches}
\end{figure}

We solve this model using DMFT and Dynamical Cluster Approximation (DCA). DMFT starts by approximating the lattice self-energy locally with that of a single site impurity model. This neglects all $\mathbf{k}$-dependence of the lattice self-energy.  DCA retains some of the momentum dependence by first cutting the Brillouin zone into patches of equal size, each patch $\mathcal{P}_\mathbf{K}$ enclosing a coarse grained momentum $\mathbf{K}$. The lattice self-energy is then approximated by a piecewise constant function over the patches and identified with that of a cluster impurity model written $\mathbf{K}$-space. In this work, we choose a minimal patching\cite{Ferrero:2009} which takes into account both the symmetries and the electron-hole pocket structure of the Brillouin zone, see Fig.~\ref{fig:BrillouinPatches}. 
One patch ($\mathcal{P}_+$) encloses the holes at $(0,0)$ and $(\pi,\pi)$ and the other patch ($\mathcal{P}_-$) encloses the electrons at $(\pi,0)$ and $(0,\pi)$.

The (cluster) impurity model is solved by continuous-time Monte-Carlo sampling of its partition function, written as a power series in the hybridization between impurity and bath (CT-HYB) \cite{GullRev:2011, Haule:2007a, Semon:2014}. This solver is well suited for strong and/or complex interactions as arising in the context of realistic material simulations. The price to pay is a complexity that scales with the dimension of the Hilbert space of the impurity. 

The 5 d-orbitals split into $e_g=\{ 3z^2 - r^2, x^2 - y^2 \}$ and $t_{2g} = \{yz, zx, xy \}$ degrees of freedom. Since the latter contribute the dominant character of the bands near the Fermi level, an idea to obtain a cluster impurity problem amenable for CT-HYB is to apply DCA only to the $t_{2g}$ orbitals, while the $e_g$ orbitals are treated within DMFT. To make this idea more specific, it is convenient to consider DMFT and DCA as approximations of the Luttinger-Ward functional\cite{LuttingerWard:1960} $\Phi_{UJ}[\mathbf{G}]$, a functional of the dressed Green function $\mathbf{G}$ which depends on the interacting part of the problem only, that is $U$ and $J$ in our case. Its derivative is the self-energy, and together with the Dyson equation
\begin{equation}
\label{equ:Dyson}
\mathbf{G}_0^{-1} - \mathbf{G}^{-1} = \mathbf{\Sigma}[\mathbf{G}]= \frac{1}{kT}\frac{\delta \Phi_{UJ}[\mathbf{G}]}{\delta \mathbf{G}},
\end{equation}
the (approximate) Luttinger-Ward functional determines hence the (approximate) solution of the problem with bare Green function $\mathbf{G}_{0}$. Diagrammatically, the Luttinger-Ward functional is the sum of all vacuum-to-vacuum skeleton diagrams, and DMFT keeps only the diagrams with support on a site. In momentum space, this corresponds to neglect conservation of momentum at the vertices, which is partially restored in DCA by conserving at least the coarse grained momentum $\mathbf{K}$. We call the corresponding functionals $\Phi^\text{loc}_{UJ}[\mathbf{G}]$ and $\Phi^\text{cl}_{UJ}[\mathbf{G}]$, respectively. In this functional formulation, the mixed DMFT-DCA treatment of the orbitals that we propose consists in approximating the lattice functional as
\begin{equation}
\label{equ:LWApprox}
\Phi_{UJ}[\mathbf{G}] = \Phi^{\text{loc}}_{UJ}[\mathbf{G}] + \Phi^{\text{cl}}_{\tilde{U}\tilde{J}}[\hat{P}_{t_{2g}}\mathbf{G}] - \Phi^\text{loc}_{\tilde{U}\tilde{J}}[\hat{P}_{t_{2g}}\mathbf{G}],
\end{equation}
where $\hat{P}_{t_{2g}}$ is the projector on $t_{2g}$ orbitals. One can think of this as a selective improvement of the diagrammatic summation by going from single site to cluster DCA for the $t_{2g}$ orbitals which is corrected by subtracting the double counting of the single site DMFT $t_{2g}$ diagrams. The use of $\tilde{U}$ and $\tilde{J}$ reflects the screening of the bare interactions by the elimination of the $e_g$ degrees of freedom in the cluster corrections. In the supplementary information, we show how the screening is determined and how the mixed DMFT-DCA scheme is solved in practice. For the sake of completeness, the solution of the DMFT equations and the impurity models are detailed as well.

\begin{figure}[h]
\centering
\includegraphics[width=1\columnwidth]{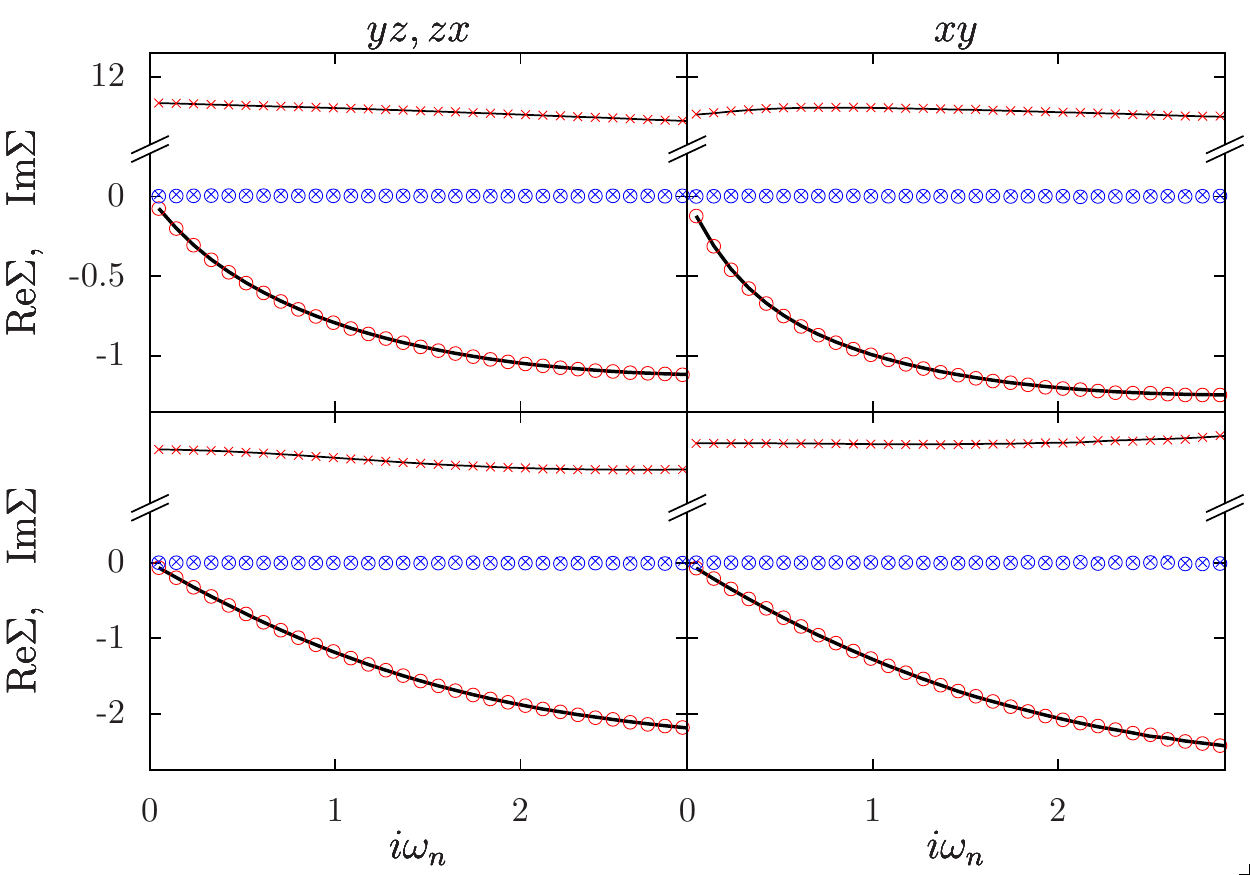}
\caption{(Color online) Comparison of the $t_{2g}$ self-energy obtained by DMFT (real/imaginary part with thin/bold lines) and DCA (local/non-local part in red/blue and real/imaginary part with crosses/circles). All self-energies are diagonal in orbital space, see supplementary information Sec.~A. The left panels show the degenerate $yz,zx$ entries and the right panel shows the $xy$ entry. The temperature is $T=174K$. In the upper panels $(U,\tilde{U})=(4.5,4.5)$ and $(J,\tilde{J})=(0.45,0.375)$ while in the lower panels $(U,\tilde{U})=(10.125,9)$ and $(J,\tilde{J})=(0,0)$. }
\label{fig:DCAvsDMFT}
\end{figure}

In the following, all energies are given in units of $eV$ and the filling is constrained to 6 electrons per $Fe$ atom. 
The upper panel in Fig.~\ref{fig:DCAvsDMFT} shows the $t_{2g}$ the self-energies obtained by DMFT and DCA at $T=174K$, $(U,\tilde{U})=(4.5,4.5)$ and $(J,\tilde{J})=(0.45,0.375)$. The DCA self-energy is shown in a ``real-space site basis" with local part $(\mathbf{\Sigma}_{\mathbf{K}=+} + \mathbf{\Sigma}_{\mathbf{K}=-})/2$  
and non-local part $(\mathbf{\Sigma}_{\mathbf{K}=+} - \mathbf{\Sigma}_{\mathbf{K}=-})/2$. The non-local self-energy is essentially zero and the local self-energy is in excellent agreement with DMFT. The quasiparticle weight is $(Z_{yz/zx},Z_{xy})=(0.4,0.3)$ and the filling of the $t_{2g}$-filling per $Fe$ atom is $N_{t_{2g}}=3.186$. To address the question  
\begin{figure}[h]
\centering
\includegraphics[width=1.0\columnwidth]{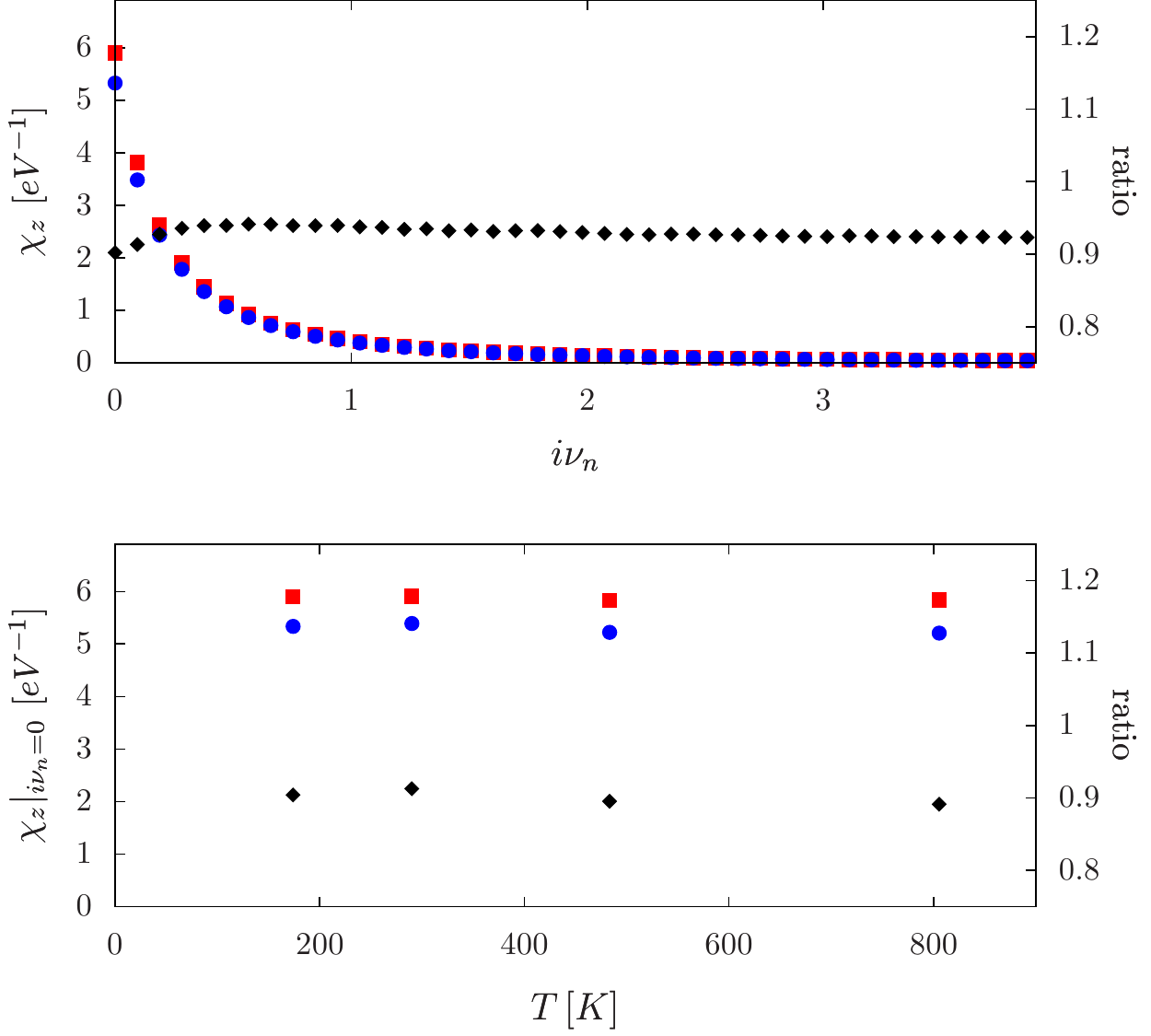}
\caption{(Color online) Impurity spin susceptibility (see Eq.~\ref{equ:susceptibility}) for DMFT (red squares) and DCA (blue circles) 
for parameters $(U,\tilde{U})=(4.5,4.5)$ and $(J,\tilde{J})=(0.45,0.375)$.
The upper panel plots the susceptibility as a function of Matsubara frequencies for $T=174K$.  The lower panel plots the susceptibility as function of the temperature for $i\nu_n=0$. 
The black diamonds display  the ratio between the DMFT and DCA spin susceptibility, and its scale is displayed in 
the right y-axis.}
\label{fig:Susc}
\end{figure}
wether this is due to the Hund's rule coupling or the orbital degeneracy, we set $J=0$ but increase $U$ in order to stay in a correlated regime, see lower panel in Fig.~\ref{fig:DCAvsDMFT}. The self-energies are local as well, $(Z_{yz/zx},Z_{xy})=(0.41,0.39)$ and $N_{t_{2g}}=2.77$. 

Another question that arises is the locality at the two particle level. To this end, we measure the impurity spin susceptibility defined as 
\begin{equation}
\label{equ:susceptibility}
\chi_{z}(i \nu_n) = \frac{1}{N_{\mathcal{P}}}\int_{0}^\beta e^{i\nu_n \tau}\langle S^{t_{2g}}_z(\tau) S^{t_{2g}}_z\rangle d\tau,
\end{equation}
where $S_z^{t_{2g}}$ is the total spin along the $z$ direction of the $t_{2g}$ degrees of freedom on the impurity, for both DMFT ($N_{\mathcal{P}}=1$) and DCA ($N_{\mathcal{P}}=2$), see Fig.~\ref{fig:Susc}. We also plot the ratio of the DCA and DMFT susceptibility which is $\approx 0.9$, meaning that even at the two particle level, our coarse graining does not show momentum space differentiation. This is very different from the cuprate case. 

\begin{figure}[h]
\centering
\includegraphics[width=1.0\columnwidth]{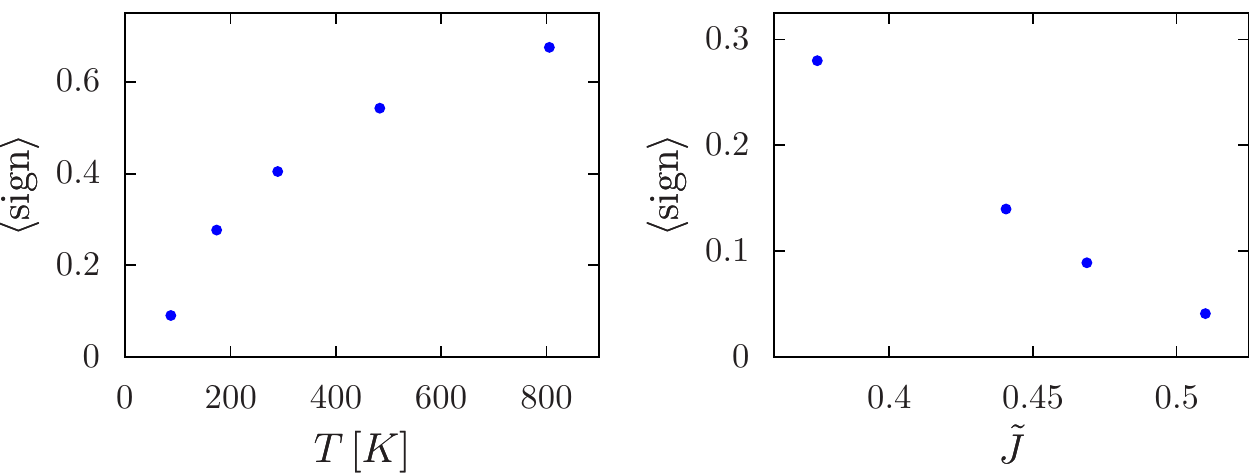}
\caption{Average sign in the CT-HYB simulations for the DCA impurity model, in the left panel for $(J,\tilde{J})=(0.45,0.375)$ as function of the temperature and in the right panel for $T=174K$ as a function of the Hund's rule coupling (the values are $(J,\tilde{J})=(0.45,0.375)$, $(0.488,0.441)$, $(0.506,0.469)$ and $(0.525,0.51)$ from left to right).  The Coulomb repulsion is $(U,\tilde{U})=(4.5,4.5)$ in both panels. The CT-HYB simulations are carried out in the $\mathbf{K}$-space single particle basis, see supplementary information Sec.~D.}
\label{fig:Sign}
\end{figure}

Fig.~\ref{fig:Sign} shows the average sign in the CT-HYB simulations for the DCA impurity model for different temperatures and Hund's rule couplings. The sign rapidly drops with increasing Hund's rule coupling. This makes cluster simulations of materials with large mass renormalizations expensive, in particular at low temperatures.


To conclude, we have demonstrated that the local approximation describes well Hunds metals, such as many iron-pnictides and chalcogenides, in their normal state. In the region of large mass renormalizations, relevant to materials such as $FeTe$, there is an onset of a severe minus sign problem.  In itself this does not prove non-locality of the self-energies, but the investigation of this region will require other impurity solvers and is outside the scope of this work.

We have solved  the same model Hamiltonian with other two site tiling of the Brillouin zone.  The results support our conclusion that the self energy is local, with little tendency towards momentum space differentiation in the parameter range explored in this paper. 
Recently a three band Hamiltonian with nearest neighbors on the square lattice and strong Hunds and Hubbard interactions was studied.\cite{Nomura:2015} 
Strong momentum space differentiation was found  for much larger values of the Hunds coupling and the Hubbard U.

Comparing our results with Ref.~\onlinecite{Nomura:2015} raises the question of what are the essential ingredients (dispersion relation, filling or interaction strength) 
needed to obtain momentum space differentiation in multi-orbital problems.
The technical advances introduced in this paper make possible the investigation of symmetry breaking phases. Future work will apply this formalism to address nematic, magnetic and superconducting order in the iron pnictides and chalcogenides. 

This work was supported by NSF. NSF-DMR1308141 (P.S. and G.K.) and NSF-DMR1405303 (K.H.).  This research used resources of the Oak Ridge Leadership Computing Facility at the Oak Ridge National Laboratory, which is supported by the Office of Science of the US Department of Energy under Contract No. DE-AC05-00OR22725.


\newpage

\setcounter{equation}{0}
\setcounter{figure}{0}
\setcounter{page}{1}
\makeatletter
\renewcommand{\theequation}{S\arabic{equation}}
\renewcommand{\thefigure}{S\arabic{figure}}
\renewcommand{\bibnumfmt}[1]{[S#1]}
\renewcommand{\citenumfont}[1]{S#1}

\section{Supplementary Information}
We begin here by writing down the equations for the the mixed DMFT-DCA scheme as defined by Eqs.~\ref{equ:Dyson} and \ref{equ:LWApprox} by means of impurity models. We then show how the effective interactions $\tilde{U}$ and $\tilde{J}$ are determined and the equations are solved in practice. Finally, we detail the impurity models.

\subsection{A. Mixed DMFT-DCA equations}
The functional derivative of Eq.~\ref{equ:LWApprox} yields the approximation 
\begin{equation}
\label{equ:SelfApprox}
\mathbf{\Sigma}(\mathbf{k})= \left (
\begin{array}{cc}
    \mathbf{\Sigma}^\text{loc}_{t_{2g}} + \mathbf{\tilde{\Sigma}}^\text{cl}_\mathbf{K} - \mathbf{\tilde{\Sigma}}^\text{loc}& 0 \\
    0 & \mathbf{\Sigma}_{e_g}^\text{loc}
\end{array}
\right )
\end{equation}
for the lattice self-energy written in $\mathbf{k}$-space, where $\mathbf{K}=+$ $(\mathbf{K}=-)$ if $\mathbf{k}$ lies in the patch $\mathcal{P}_+$ ($\mathcal{P}_-$). The self-energies on the right hand side of Eq.~\ref{equ:SelfApprox} are identified with those of impurity models as follows:

\begin{itemize}
\item[(i)] $\mathbf{\Sigma}^\text{loc}$, a diagonal $5\times 5$ matrix in $d$-shell orbital space with components $\mathbf{\Sigma}^\text{loc}_{t_{2g}}$ and $\mathbf{\Sigma}^\text{loc}_{e_g}$, is the self-energy of a single site $d$-shell impurity model with interactions $U$ and $J$. 
\item[(ii)] $\mathbf{\tilde{\Sigma}}^\text{loc}$, a diagonal $3\times 3$ matrix in $t_{2g}$ orbital space, is the self-energy of a single site $t_{2g}$-orbital impurity model with effective interactions $\tilde{U}$ and $\tilde{J}$. 
\item[(iii)] $\mathbf{\tilde{\Sigma}}^\text{cl}_\mathbf{K}$, a diagonal $3\times 3$ matrix in $t_{2g}$ orbital space for each $\mathbf{K}$, is the self-energy of a two-site $t_{2g}$-orbital cluster impurity model with effective interactions $\tilde{U}$ and $\tilde{J}$. 
\end{itemize}

The non-interacting part of these impurity models is encapsulated in the Weiss-Fields $\mathbf{G}_0^\text{loc}$, $\mathbf{\tilde{G}}_0^\text{loc}$ and $\mathbf{\tilde{G}}^\text{cl}_{0\mathbf{K}}$, which relate the self-energies with the interacting Greens functions $\mathbf{G}^\text{loc}$, $\mathbf{\tilde{G}}^\text{loc}$ and $\mathbf{\tilde{G}}^\text{cl}_\mathbf{K}$ through the Dyson equations 
\begin{subequations}
\label{equ:ImpurityDyson}
\begin{align}
(\mathbf{G}^\text{loc})^{-1}&=(\mathbf{G}_0^\text{loc})^{-1} -  \mathbf{\Sigma}^\text{loc}\\ 
(\mathbf{\tilde{G}}^\text{loc})^{-1}&=(\mathbf{\tilde{G}}_{0}^\text{loc})^{-1} - \mathbf{\tilde{\Sigma}}^\text{loc}\\
(\mathbf{\tilde{G}}^\text{cl}_\mathbf{K})^{-1}&=(\mathbf{\tilde{G}}_{0\mathbf{K}}^\text{cl})^{-1} - \mathbf{\tilde{\Sigma}}^\text{cl}_\mathbf{K}.
\end{align}   
\end{subequations}
Eq.~\ref{equ:SelfApprox} yields the approximate lattice Green function 
\begin{equation}
\label{equ:LatticeDyson}
\mathbf{G}^{-1}(\mathbf{k}) = \mathbf{G}_0^{-1}(\mathbf{k}) - \mathbf{\Sigma}(\mathbf{k}),
\end{equation} 
where $\mathbf{G}_0^{-1}(i\omega_n,\mathbf{k}) = i\omega_n + \mu - \mathbf{h}_0(\mathbf{k})$ is the bare lattice Green function. The DMFT and DCA approximations of the Luttinger-Ward functional then require 
\begin{subequations}
\label{equ:SelfConsistency}
\begin{align}
\mathbf{G}^\text{loc}&=\frac{1}{|\text{BZ}|}\int_\text{BZ} d\mathbf{k} \mathbf{G}(\mathbf{k})\\
\mathbf{\tilde{G}}^\text{loc} &= \frac{1}{|\text{BZ}|} \int_{\text{BZ}} d\mathbf{k}\hat{P}_{t_{2g}}\mathbf{G}(\mathbf{k})\\
\mathbf{\tilde{G}}_\mathbf{K}^\text{cl} &= \frac{1}{|\mathcal{P}_\mathbf{K}|}\int_{\mathcal{P}_\mathbf{K}} d\mathbf{k} \hat{P}_{t_{2g}} \mathbf{G}(\mathbf{k}).
\end{align}
\end{subequations}
Fixing the chemical potential by imposing 6 electrons per atom, above equations determine the Weiss-Fields and hereby the solution of the mixed DMFT-DCA scheme. The interactions $U$ and $J$ are taken as external parameters, and what remains to be determined are the effective interactions $\tilde{U}$ and $\tilde{J}$ which take into account the screening of the $t_{2g}$ degrees of freedom in the cluster corrections. 

Notice here that, in the normal phase, the DMFT self-energies $\mathbf{\Sigma}^\text{loc}$ and $\mathbf{\tilde{\Sigma}}^\text{loc}$ (and also $\mathbf{G}_0^\text{loc}$, $\mathbf{\tilde{G}}_0^\text{loc}$, $\mathbf{G}^\text{loc}$ and $\mathbf{\tilde{G}}^\text{loc}$) are diagonal in the orbital space. This comes from the $D_{2d}$ point symmetry group of an $Fe$ atom. Furthermore, the patches are invariant under this symmetry group, so that the DCA self-energy $\mathbf{\tilde{\Sigma}}_\mathbf{K}^\text{cl}$ (and also $\mathbf{\tilde{G}}_{0\mathbf{K}}^\text{cl}$ and $\mathbf{\tilde{G}}_{\mathbf{K}}^\text{cl}$) is diagonal in the orbital space as well.

\subsection{B. Effective interactions}
\begin{figure}[h]
\centering
\includegraphics[width=\columnwidth]{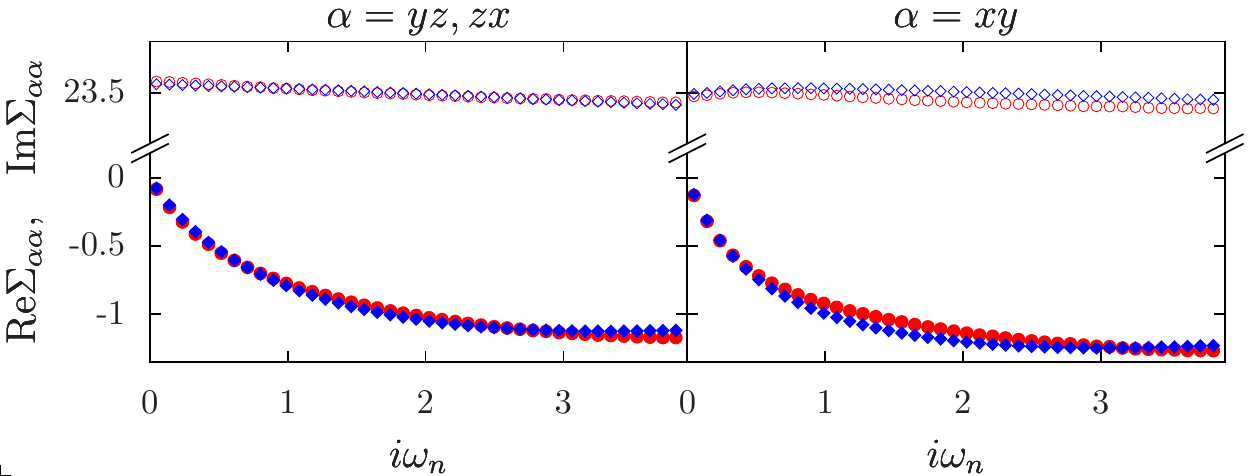}
\caption{(Color online) Self-energies (real/imaginary part with open/filled symbols) of the $t_{2g}$ orbitals obtained with DMFT for the model with interactions applied to all $d$-shell orbitals (red circles) and for the effective $t_{2g}$ model (blue diamonds). In the latter case, the Hartree-Fock constant $\Sigma^{\text{HF}}$ is added. The parameters are $T=174K$, $(U,\tilde{U})=(4.5,4.5)$ and $(J,\tilde{J})=(0.45,0.375)$.}
\label{fig:t2gApproximation}
\end{figure}

To determine the effective interactions, we define an effective problem where correlations are applied only to the $t_{2g}$ orbitals. These effective correlations, which are identified with $\tilde{U}$ and $\tilde{J}$, are then determined by requiring that this model reproduces at low energies the results of the five band calculation (with correlations $U$ and $J$), \textit{when both models are solved via single site DMFT}. We use the following algorithm:
\begin{itemize}
\item[(i)] The five band model is solved with single site DMFT for a filling of 6 $d$-shell electrons per $Fe$ atom and interactions $U$ and $J$. This yields a local lattice self-energy $\mathbf{\Sigma}^{\text{loc}}$ (with components $\mathbf{\Sigma}_{e_g}^\text{loc}$ and $\mathbf{\Sigma}_{t_{2g}}^\text{loc}$), a filling of the $t_{2g}$ orbitals and a chemical potential.
\item[(ii)] We determine the low energy model by defining the effective bare lattice propagator
\begin{equation}
\label{equ:effective_propagator}
\mathbf{\tilde{G}}_0^{-1} := \mathbf{G}_0^{-1}-\left (
\begin{array}{cc}
    \Sigma^{\text{HF}}\cdot \mathbf{1}_{t_{2g}}& 0 \\
    0 & \mathbf{\Sigma}_{e_g}^\text{loc}
\end{array}
\right ).
\end{equation}
Here, $\mathbf{\Sigma}^\text{loc}_{e_g}$ is the $e_g$ part of the self-energy obtained in (i), and $\Sigma^{\text{HF}}$  (which can be thought as an average Hartree-Fock contribution to the $t_{2g}$ self-energy coming from the $e_g$ orbitals) will be determined in the next step. The chemical potential is fixed to the value obtained in (i).
\item[(iii)] To determine  $\Sigma^{\text{HF}}$,  $\tilde{U}$ and $\tilde{J}$, we solve the problem with propagator Eq.~\ref{equ:effective_propagator} and the effective interactions applied to the $t_{2g}$ orbitals with single-site DMFT. The resulting self-energy is denoted by $\mathbf{\tilde{\Sigma}}^\text{loc}$.
 $\Sigma^{\text{HF}}$ is determined by requiring that the $t_{2g}$ filling is the same as in (i). 
Requiring that  $\mathbf{\tilde{\Sigma}}^\text{loc}+\Sigma^{\text{HF}}$ matches  $\mathbf{\Sigma}_{t_{2g}}^\text{loc} $ at the  lowest Matsubara frequencies determines the effective interactions $\tilde{U}$ and $\tilde{J}$. 
\end{itemize}
It is remarkable that these requirements give us very good matching of the self-energies at all energies, as shown in Fig.~\ref{fig:t2gApproximation}. 

\subsection{C. Solving the mixed DMFT-DCA equations and the DMFT equation}
The good agreement of the self-energies in Fig.~\ref{fig:t2gApproximation} suggests to solve the mixed DMFT-DCA scheme in a simplified manner. Instead of simultaneously solving the three impurity models in Sec.~A, we just apply DCA to the effective $t_{2g}$ model used to determine the screened interactions $\tilde{U}$ and $\tilde{J}$ . $\Sigma^{\text{HF}}$ is slightly readjusted to preserve the $t_{2g}$ filling found in Sec.~B (i).

This simplified solution is justified \textit{if the cluster corrections to local quantities is small}. Indeed, in this case we can start by ignoring the cluster corrections when solving the mixed DMFT-DCA scheme and solve the model with DMFT (which corresponds to step (i) in Sec.~B). We then solve the model with cluster corrections (where $\tilde{U}$ and $\tilde{J}$ have been determined as discussed in Sec.~B), keeping however the $e_g$ self-energy $\mathbf{\Sigma}_{e_g}^\text{loc}$ and the chemical potential obtained without cluster corrections fixed. Further, the contribution to the $t_{2g}$ self-energy from $\mathbf{\Sigma}^\text{loc}_{t_{2g}} - \mathbf{\tilde{\Sigma}}^\text{loc}$ is replaced by a constant proportional to the identity, which is justified by Fig.~\ref{fig:t2gApproximation}. Choosing this constant $\Sigma^{\text{HF}}$ to preserve the $t_{2g}$ filling found in Sec.~B (i), this amounts just to solve the effective $t_{2g}$ model with DCA as mentioned above.
Compared to the exact solution of the DMFT-DCA scheme, this simplified solution avoids stability issues and guaranties causality (c.f. nested cluster schemes in Ref.~\onlinecite{Biroli:2004}). 

When comparing  results from the mixed DMFT-DCA scheme with DMFT results, the latter is applied to the effective $t_{2g}$ problem for the sake of coherence. The DMFT self-energy is thus $\mathbf{\tilde{\Sigma}}^\text{loc}$ from Sec.~B (iii), while the mixed DMFT-DCA self-energy is $\mathbf{\Sigma}_\mathbf{K}^\text{cl}$, obtained in the above approximation.

\subsection{D. Impurity Models}

The aim here is to write down the action for the impurity models in Sec.~A. To this end, we begin by detailing the interaction used in this work.

With respect to the $d$-shell single particle basis $|\sigma m \rangle$, where $\sigma$ is the spin and the angular part is encapsulated in the spherical harmonics $Y_{l=2}^m$, the local interaction 
\begin{equation}
\label{equ:interaction}
\hat{V} = \frac{1}{2}\sum_{\sigma \sigma'}\sum_{\{m_i\}}V_{m_1m_2m_3m_4}c^\dagger_{\sigma m_1}c^\dagger_{\sigma' m_2}c_{\sigma' m_3}c_{\sigma m_4}
\end{equation}
is given by the tensor 
\begin{equation}
\label{equ:tensor}
\begin{split}
&V_{m_1m_2m_3m_4} = \sum_{k=0,2,4} \frac{4\pi}{2k+1}F^k  \\
&\quad \times \sum_{m=-k}^k \langle Y_{2}^{m_1} | Y_k^{m*} | Y_2^{m_4}\rangle \langle Y_{2}^{m_2} | Y_k^m | Y_2^{m_3}\rangle.
\end{split}
\end{equation} 
The Slater-Condon parameters $F^0$, $F^2$ and $F^4$ encapsulate both the radial part of the single particle basis (which is the same for all $|\sigma m\rangle$) and the interaction. 
In this work we use $F^0=U$, $F^2=14\cdot J /1.625$ and $F^4=0.625 \cdot F^2$, where $U$ is the Coulomb repulsion and $J$ the Hund's rule coupling. 

In solids, it is more convenient to work in the basis of real spherical harmonics $|\sigma \alpha\rangle$ with $\alpha \in \{yz,zx,xy,3z^2-r^2,x^2-y^2\}$, and we denote the corresponding creation operators by $d_{\sigma \alpha}^\dagger$. We slightly simplify the interaction tensor $V_{\alpha_1\alpha_2\alpha_3\alpha_4}$ in this basis by setting all elements which are not of the form $V_{\alpha\alpha\alpha'\alpha'}$,$V_{\alpha\alpha'\alpha'\alpha}$ or $V_{\alpha\alpha'\alpha\alpha'}$ to zero. While this truncation preserves the spin $SU(2)$ invariance of the interaction Eqs.~\ref{equ:interaction} and \ref{equ:tensor}, the orbital $SO(3)$ invariance is lifted. However, the truncated interaction is still $D_{2d}$ invariant and the crystal fields in the present case lift the $SO(3)$ degeneracy anyway.

In the basis of real spherical harmonics, the action for the single-site $d$-shell impurity model Sec.~A (i) reads
\begin{equation}
\label{equ:DMFTAction}
\begin{split}
S =& -\sum_\sigma \sum_\alpha \iint_0^\beta  d^\dagger_{\sigma \alpha}(\tau)G^{-1}_{0\alpha \alpha }(\tau - \tau')d_{\sigma \alpha}(\tau')d\tau d\tau' \\
&+\frac{1}{2}\sum_{\sigma\sigma'}\sum_{\{\alpha_i\}} V^{UJ}_{\alpha_1\alpha_2\alpha_3\alpha_4}\int_0^\beta d_{\sigma \alpha_1}^\dagger(\tau) d_{\sigma'\alpha_2}^\dagger(\tau)  \\
&\times d_{\sigma'\alpha_3}(\tau)d_{\sigma \alpha_4}(\tau) d\tau,
\end{split}
\end{equation}
where the superscript of the interaction tensor indicates that $U$ and $J$ enter the Slater-Condon parameters. Restricting in the action Eq.~\ref{equ:DMFTAction} the orbital sums to $t_{2g}$ orbitals and replacing $U$, $J$ and $\mathbf{G}_0$ by the effective $\tilde{U}$, $\tilde{J}$ and $\mathbf{\tilde{G}}_0$ respectively yields the single-site $t_{2g}$ impurity model Sec.~A (ii). 

The non-interacting part of the cluster impurity model action Sec.~A (iii) reads 
\begin{equation}
\begin{split}
S_0 = -\sum_\sigma \sum_{\mathbf{K}=\pm}\sum_{\alpha \in t_{2g}}&\iint_0^\beta  d^\dagger_{\sigma\alpha\mathbf{K}}(\tau)\tilde{G}^{-1}_{0\mathbf{K}\alpha \alpha }(\tau - \tau')\\
&\times d_{\sigma \alpha \mathbf{K}}(\tau')d\tau d\tau',
\end{split}
\end{equation} 
where $d_{\sigma \alpha \mathbf{K}}^\dagger$ creates an electron with spin $\sigma$ and coarse grained momentum $\mathbf{K}$ in the orbital $\alpha$. 
The interacting part, written in a ``real-space site basis" $d_{\alpha\sigma 1}:=(d_{\sigma\alpha +} + d_{\sigma \alpha -})/\sqrt 2$ and $d_{\alpha\sigma 2}:=(d_{\sigma\alpha +} - d_{\sigma\alpha -})/\sqrt 2$, reads
\begin{equation}
\label{equ:DCAinteraction}
\begin{split}
S_I=&\frac{1}{2}\sum_{\sigma\sigma'}\sum_{a=1,2}\sum_{\lbrace \alpha_i\rbrace \in t_{2g}}V^{\tilde{U}\tilde{J}}_{\alpha_1 \alpha_2 \alpha_3 \alpha_4}  \\
&\times \int_0^\beta d^\dagger_{\sigma \alpha_1  a}(\tau)d^\dagger_{\sigma' \alpha_2  a} (\tau)d_{\sigma' \alpha_3 a}(\tau)d_{\sigma \alpha_4 a}(\tau) d\tau.
\end{split}
\end{equation}
For the CT-HYB simulations, the $\mathbf{K}$-space single particle basis $d_{\sigma \mathbf{K} \alpha}^\dagger$ is used.


\begin{thebibliography}{54}%
\makeatletter
\providecommand \@ifxundefined [1]{%
 \@ifx{#1\undefined}
}%
\providecommand \@ifnum [1]{%
 \ifnum #1\expandafter \@firstoftwo
 \else \expandafter \@secondoftwo
 \fi
}%
\providecommand \@ifx [1]{%
 \ifx #1\expandafter \@firstoftwo
 \else \expandafter \@secondoftwo
 \fi
}%
\providecommand \natexlab [1]{#1}%
\providecommand \enquote  [1]{``#1''}%
\providecommand \bibnamefont  [1]{#1}%
\providecommand \bibfnamefont [1]{#1}%
\providecommand \citenamefont [1]{#1}%
\providecommand \href@noop [0]{\@secondoftwo}%
\providecommand \href [0]{\begingroup \@sanitize@url \@href}%
\providecommand \@href[1]{\@@startlink{#1}\@@href}%
\providecommand \@@href[1]{\endgroup#1\@@endlink}%
\providecommand \@sanitize@url [0]{\catcode `\\12\catcode `\$12\catcode
  `\&12\catcode `\#12\catcode `\^12\catcode `\_12\catcode `\%12\relax}%
\providecommand \@@startlink[1]{}%
\providecommand \@@endlink[0]{}%
\providecommand \url  [0]{\begingroup\@sanitize@url \@url }%
\providecommand \@url [1]{\endgroup\@href {#1}{\urlprefix }}%
\providecommand \urlprefix  [0]{URL }%
\providecommand \Eprint [0]{\href }%
\providecommand \doibase [0]{http://dx.doi.org/}%
\providecommand \selectlanguage [0]{\@gobble}%
\providecommand \bibinfo  [0]{\@secondoftwo}%
\providecommand \bibfield  [0]{\@secondoftwo}%
\providecommand \translation [1]{[#1]}%
\providecommand \BibitemOpen [0]{}%
\providecommand \bibitemStop [0]{}%
\providecommand \bibitemNoStop [0]{.\EOS\space}%
\providecommand \EOS [0]{\spacefactor3000\relax}%
\providecommand \BibitemShut  [1]{\csname bibitem#1\endcsname}%
\let\auto@bib@innerbib\@empty
\bibitem [{\citenamefont {Kamihara}\ \emph {et~al.}(2006)\citenamefont
  {Kamihara}, \citenamefont {Hiramatsu}, \citenamefont {Hirano}, \citenamefont
  {Kawamura}, \citenamefont {Yanagi}, \citenamefont {Kamiya},\ and\
  \citenamefont {Hosono}}]{Hosono:2006}%
  \BibitemOpen
  \bibfield  {author} {\bibinfo {author} {\bibfnamefont {Y.}~\bibnamefont
  {Kamihara}}, \bibinfo {author} {\bibfnamefont {H.}~\bibnamefont {Hiramatsu}},
  \bibinfo {author} {\bibfnamefont {M.}~\bibnamefont {Hirano}}, \bibinfo
  {author} {\bibfnamefont {R.}~\bibnamefont {Kawamura}}, \bibinfo {author}
  {\bibfnamefont {H.}~\bibnamefont {Yanagi}}, \bibinfo {author} {\bibfnamefont
  {T.}~\bibnamefont {Kamiya}}, \ and\ \bibinfo {author} {\bibfnamefont
  {H.}~\bibnamefont {Hosono}},\ }\href {\doibase 10.1021/ja063355c} {\bibfield
  {journal} {\bibinfo  {journal} {Journal of the American Chemical Society}\
  }\textbf {\bibinfo {volume} {128}},\ \bibinfo {pages} {10012} (\bibinfo
  {year} {2006})}\BibitemShut {NoStop}%
\bibitem [{\citenamefont {{Chubukov}}\ and\ \citenamefont
  {{Hirschfeld}}(2014)}]{ChubukovHirschfeld:2014}%
  \BibitemOpen
  \bibfield  {author} {\bibinfo {author} {\bibfnamefont {A.~V.}\ \bibnamefont
  {{Chubukov}}}\ and\ \bibinfo {author} {\bibfnamefont {P.~J.}\ \bibnamefont
  {{Hirschfeld}}},\ }\href@noop {} {\bibfield  {journal} {\bibinfo  {journal}
  {ArXiv e-prints}\ } (\bibinfo {year} {2014})},\ \Eprint
  {http://arxiv.org/abs/1412.7104} {arXiv:1412.7104 [cond-mat.supr-con]}
  \BibitemShut {NoStop}%
\bibitem [{\citenamefont {Haule}\ and\ \citenamefont
  {Kotliar}(2009)}]{HauleKotliar:2009}%
  \BibitemOpen
  \bibfield  {author} {\bibinfo {author} {\bibfnamefont {K.}~\bibnamefont
  {Haule}}\ and\ \bibinfo {author} {\bibfnamefont {G.}~\bibnamefont
  {Kotliar}},\ }\href {http://stacks.iop.org/1367-2630/11/i=2/a=025021}
  {\bibfield  {journal} {\bibinfo  {journal} {New Journal of Physics}\ }\textbf
  {\bibinfo {volume} {11}},\ \bibinfo {pages} {025021} (\bibinfo {year}
  {2009})}\BibitemShut {NoStop}%
\bibitem [{\citenamefont {Yin}\ \emph {et~al.}(2011)\citenamefont {Yin},
  \citenamefont {Haule},\ and\ \citenamefont {Kotliar}}]{YinHauleKotliar:2011}%
  \BibitemOpen
  \bibfield  {author} {\bibinfo {author} {\bibfnamefont {Z.~P.}\ \bibnamefont
  {Yin}}, \bibinfo {author} {\bibfnamefont {K.}~\bibnamefont {Haule}}, \ and\
  \bibinfo {author} {\bibfnamefont {G.}~\bibnamefont {Kotliar}},\ }\href
  {\doibase 10.1038/nphys1923} {\bibfield  {journal} {\bibinfo  {journal}
  {Nature Physics}\ }\textbf {\bibinfo {volume} {advance online publication}}
  (\bibinfo {year} {2011}),\ 10.1038/nphys1923}\BibitemShut {NoStop}%
\bibitem [{\citenamefont {Mravlje}\ \emph {et~al.}(2011)\citenamefont
  {Mravlje}, \citenamefont {Aichhorn}, \citenamefont {Miyake}, \citenamefont
  {Haule}, \citenamefont {Kotliar},\ and\ \citenamefont
  {Georges}}]{Mravlje:2011}%
  \BibitemOpen
  \bibfield  {author} {\bibinfo {author} {\bibfnamefont {J.}~\bibnamefont
  {Mravlje}}, \bibinfo {author} {\bibfnamefont {M.}~\bibnamefont {Aichhorn}},
  \bibinfo {author} {\bibfnamefont {T.}~\bibnamefont {Miyake}}, \bibinfo
  {author} {\bibfnamefont {K.}~\bibnamefont {Haule}}, \bibinfo {author}
  {\bibfnamefont {G.}~\bibnamefont {Kotliar}}, \ and\ \bibinfo {author}
  {\bibfnamefont {A.}~\bibnamefont {Georges}},\ }\href {\doibase
  10.1103/PhysRevLett.106.096401} {\bibfield  {journal} {\bibinfo  {journal}
  {Phys. Rev. Lett.}\ }\textbf {\bibinfo {volume} {106}},\ \bibinfo {pages}
  {096401} (\bibinfo {year} {2011})}\BibitemShut {NoStop}%
\bibitem [{\citenamefont {Georges}\ \emph {et~al.}(2013)\citenamefont
  {Georges}, \citenamefont {Medici},\ and\ \citenamefont
  {Mravlje}}]{Antoine:2013}%
  \BibitemOpen
  \bibfield  {author} {\bibinfo {author} {\bibfnamefont {A.}~\bibnamefont
  {Georges}}, \bibinfo {author} {\bibfnamefont {L.~d.}\ \bibnamefont {Medici}},
  \ and\ \bibinfo {author} {\bibfnamefont {J.}~\bibnamefont {Mravlje}},\ }\href
  {\doibase 10.1146/annurev-conmatphys-020911-125045} {\bibfield  {journal}
  {\bibinfo  {journal} {Annual Review of Condensed Matter Physics}\ }\textbf
  {\bibinfo {volume} {4}},\ \bibinfo {pages} {137} (\bibinfo {year}
  {2013})}\BibitemShut {NoStop}%
\bibitem [{\citenamefont {Georges}\ \emph {et~al.}(1996)\citenamefont
  {Georges}, \citenamefont {Kotliar}, \citenamefont {Krauth},\ and\
  \citenamefont {Rozenberg}}]{Georges:1996}%
  \BibitemOpen
  \bibfield  {author} {\bibinfo {author} {\bibfnamefont {A.}~\bibnamefont
  {Georges}}, \bibinfo {author} {\bibfnamefont {G.}~\bibnamefont {Kotliar}},
  \bibinfo {author} {\bibfnamefont {W.}~\bibnamefont {Krauth}}, \ and\ \bibinfo
  {author} {\bibfnamefont {M.~J.}\ \bibnamefont {Rozenberg}},\ }\href {\doibase
  10.1103/RevModPhys.68.13} {\bibfield  {journal} {\bibinfo  {journal} {Rev.
  Mod. Phys.}\ }\textbf {\bibinfo {volume} {68}},\ \bibinfo {pages} {13}
  (\bibinfo {year} {1996})}\BibitemShut {NoStop}%
\bibitem [{\citenamefont {Kotliar}\ \emph {et~al.}(2001)\citenamefont
  {Kotliar}, \citenamefont {Savrasov}, \citenamefont {P\'alsson},\ and\
  \citenamefont {Biroli}}]{Kotliar:2001}%
  \BibitemOpen
  \bibfield  {author} {\bibinfo {author} {\bibfnamefont {G.}~\bibnamefont
  {Kotliar}}, \bibinfo {author} {\bibfnamefont {S.~Y.}\ \bibnamefont
  {Savrasov}}, \bibinfo {author} {\bibfnamefont {G.}~\bibnamefont {P\'alsson}},
  \ and\ \bibinfo {author} {\bibfnamefont {G.}~\bibnamefont {Biroli}},\ }\href
  {\doibase 10.1103/PhysRevLett.87.186401} {\bibfield  {journal} {\bibinfo
  {journal} {Phys. Rev. Lett.}\ }\textbf {\bibinfo {volume} {87}},\ \bibinfo
  {pages} {186401} (\bibinfo {year} {2001})}\BibitemShut {NoStop}%
\bibitem [{\citenamefont {Maier}\ \emph
  {et~al.}(2005{\natexlab{a}})\citenamefont {Maier}, \citenamefont {Jarrell},
  \citenamefont {Pruschke},\ and\ \citenamefont {Hettler}}]{MaierRev:2005}%
  \BibitemOpen
  \bibfield  {author} {\bibinfo {author} {\bibfnamefont {T.}~\bibnamefont
  {Maier}}, \bibinfo {author} {\bibfnamefont {M.}~\bibnamefont {Jarrell}},
  \bibinfo {author} {\bibfnamefont {T.}~\bibnamefont {Pruschke}}, \ and\
  \bibinfo {author} {\bibfnamefont {M.~H.}\ \bibnamefont {Hettler}},\ }\href
  {\doibase 10.1103/RevModPhys.77.1027} {\bibfield  {journal} {\bibinfo
  {journal} {Rev. Mod. Phys.}\ }\textbf {\bibinfo {volume} {77}},\ \bibinfo
  {pages} {1027} (\bibinfo {year} {2005}{\natexlab{a}})}\BibitemShut {NoStop}%
\bibitem [{\citenamefont {Deng}\ \emph {et~al.}(2014)\citenamefont {Deng},
  \citenamefont {Sternbach}, \citenamefont {Haule}, \citenamefont {Basov},\
  and\ \citenamefont {Kotliar}}]{Deng:2014}%
  \BibitemOpen
  \bibfield  {author} {\bibinfo {author} {\bibfnamefont {X.}~\bibnamefont
  {Deng}}, \bibinfo {author} {\bibfnamefont {A.}~\bibnamefont {Sternbach}},
  \bibinfo {author} {\bibfnamefont {K.}~\bibnamefont {Haule}}, \bibinfo
  {author} {\bibfnamefont {D.~N.}\ \bibnamefont {Basov}}, \ and\ \bibinfo
  {author} {\bibfnamefont {G.}~\bibnamefont {Kotliar}},\ }\href {\doibase
  10.1103/PhysRevLett.113.246404} {\bibfield  {journal} {\bibinfo  {journal}
  {Phys. Rev. Lett.}\ }\textbf {\bibinfo {volume} {113}},\ \bibinfo {pages}
  {246404} (\bibinfo {year} {2014})}\BibitemShut {NoStop}%
\bibitem [{\citenamefont {Huscroft}\ \emph {et~al.}(2001)\citenamefont
  {Huscroft}, \citenamefont {Jarrell}, \citenamefont {Maier}, \citenamefont
  {Moukouri},\ and\ \citenamefont {Tahvildarzadeh}}]{Huscroft:2000}%
  \BibitemOpen
  \bibfield  {author} {\bibinfo {author} {\bibfnamefont {C.}~\bibnamefont
  {Huscroft}}, \bibinfo {author} {\bibfnamefont {M.}~\bibnamefont {Jarrell}},
  \bibinfo {author} {\bibfnamefont {T.}~\bibnamefont {Maier}}, \bibinfo
  {author} {\bibfnamefont {S.}~\bibnamefont {Moukouri}}, \ and\ \bibinfo
  {author} {\bibfnamefont {A.~N.}\ \bibnamefont {Tahvildarzadeh}},\ }\href
  {\doibase 10.1103/PhysRevLett.86.139} {\bibfield  {journal} {\bibinfo
  {journal} {Phys. Rev. Lett.}\ }\textbf {\bibinfo {volume} {86}},\ \bibinfo
  {pages} {139} (\bibinfo {year} {2001})}\BibitemShut {NoStop}%
\bibitem [{\citenamefont {Lichtenstein}\ and\ \citenamefont
  {Katsnelson}(2000)}]{Lichtenstein:2000}%
  \BibitemOpen
  \bibfield  {author} {\bibinfo {author} {\bibfnamefont {A.~I.}\ \bibnamefont
  {Lichtenstein}}\ and\ \bibinfo {author} {\bibfnamefont {M.~I.}\ \bibnamefont
  {Katsnelson}},\ }\href {\doibase 10.1103/PhysRevB.62.R9283} {\bibfield
  {journal} {\bibinfo  {journal} {Phys. Rev. B}\ }\textbf {\bibinfo {volume}
  {62}},\ \bibinfo {pages} {R9283} (\bibinfo {year} {2000})}\BibitemShut
  {NoStop}%
\bibitem [{\citenamefont {Jarrell}\ \emph
  {et~al.}(2001{\natexlab{a}})\citenamefont {Jarrell}, \citenamefont {Maier},
  \citenamefont {Hettler},\ and\ \citenamefont
  {Tahvildarzadeh}}]{Jarrell:2001a}%
  \BibitemOpen
  \bibfield  {author} {\bibinfo {author} {\bibfnamefont {M.}~\bibnamefont
  {Jarrell}}, \bibinfo {author} {\bibfnamefont {T.}~\bibnamefont {Maier}},
  \bibinfo {author} {\bibfnamefont {M.~H.}\ \bibnamefont {Hettler}}, \ and\
  \bibinfo {author} {\bibfnamefont {A.~N.}\ \bibnamefont {Tahvildarzadeh}},\
  }\href {http://stacks.iop.org/0295-5075/56/i=4/a=563} {\bibfield  {journal}
  {\bibinfo  {journal} {EPL (Europhysics Letters)}\ }\textbf {\bibinfo {volume}
  {56}},\ \bibinfo {pages} {563} (\bibinfo {year}
  {2001}{\natexlab{a}})}\BibitemShut {NoStop}%
\bibitem [{\citenamefont {Jarrell}\ \emph
  {et~al.}(2001{\natexlab{b}})\citenamefont {Jarrell}, \citenamefont {Maier},
  \citenamefont {Huscroft},\ and\ \citenamefont {Moukouri}}]{Jarrell:2001b}%
  \BibitemOpen
  \bibfield  {author} {\bibinfo {author} {\bibfnamefont {M.}~\bibnamefont
  {Jarrell}}, \bibinfo {author} {\bibfnamefont {T.}~\bibnamefont {Maier}},
  \bibinfo {author} {\bibfnamefont {C.}~\bibnamefont {Huscroft}}, \ and\
  \bibinfo {author} {\bibfnamefont {S.}~\bibnamefont {Moukouri}},\ }\href
  {\doibase 10.1103/PhysRevB.64.195130} {\bibfield  {journal} {\bibinfo
  {journal} {Phys. Rev. B}\ }\textbf {\bibinfo {volume} {64}},\ \bibinfo
  {pages} {195130} (\bibinfo {year} {2001}{\natexlab{b}})}\BibitemShut
  {NoStop}%
\bibitem [{\citenamefont {Haule}\ \emph {et~al.}(2003)\citenamefont {Haule},
  \citenamefont {Rosch}, \citenamefont {Kroha},\ and\ \citenamefont
  {W\"olfle}}]{Haule:2003}%
  \BibitemOpen
  \bibfield  {author} {\bibinfo {author} {\bibfnamefont {K.}~\bibnamefont
  {Haule}}, \bibinfo {author} {\bibfnamefont {A.}~\bibnamefont {Rosch}},
  \bibinfo {author} {\bibfnamefont {J.}~\bibnamefont {Kroha}}, \ and\ \bibinfo
  {author} {\bibfnamefont {P.}~\bibnamefont {W\"olfle}},\ }\href {\doibase
  10.1103/PhysRevB.68.155119} {\bibfield  {journal} {\bibinfo  {journal} {Phys.
  Rev. B}\ }\textbf {\bibinfo {volume} {68}},\ \bibinfo {pages} {155119}
  (\bibinfo {year} {2003})}\BibitemShut {NoStop}%
\bibitem [{\citenamefont {Parcollet}\ \emph {et~al.}(2004)\citenamefont
  {Parcollet}, \citenamefont {Biroli},\ and\ \citenamefont
  {Kotliar}}]{Parcollet:2004}%
  \BibitemOpen
  \bibfield  {author} {\bibinfo {author} {\bibfnamefont {O.}~\bibnamefont
  {Parcollet}}, \bibinfo {author} {\bibfnamefont {G.}~\bibnamefont {Biroli}}, \
  and\ \bibinfo {author} {\bibfnamefont {G.}~\bibnamefont {Kotliar}},\ }\href
  {\doibase 10.1103/PhysRevLett.92.226402} {\bibfield  {journal} {\bibinfo
  {journal} {Phys. Rev. Lett.}\ }\textbf {\bibinfo {volume} {92}},\ \bibinfo
  {pages} {226402} (\bibinfo {year} {2004})}\BibitemShut {NoStop}%
\bibitem [{\citenamefont {Carter}\ and\ \citenamefont
  {Schofield}(2004)}]{Carter:2004}%
  \BibitemOpen
  \bibfield  {author} {\bibinfo {author} {\bibfnamefont {E.~C.}\ \bibnamefont
  {Carter}}\ and\ \bibinfo {author} {\bibfnamefont {A.~J.}\ \bibnamefont
  {Schofield}},\ }\href {\doibase 10.1103/PhysRevB.70.045107} {\bibfield
  {journal} {\bibinfo  {journal} {Phys. Rev. B}\ }\textbf {\bibinfo {volume}
  {70}},\ \bibinfo {pages} {045107} (\bibinfo {year} {2004})}\BibitemShut
  {NoStop}%
\bibitem [{\citenamefont {Civelli}\ \emph {et~al.}(2005)\citenamefont
  {Civelli}, \citenamefont {Capone}, \citenamefont {Kancharla}, \citenamefont
  {Parcollet},\ and\ \citenamefont {Kotliar}}]{Civelli:2005}%
  \BibitemOpen
  \bibfield  {author} {\bibinfo {author} {\bibfnamefont {M.}~\bibnamefont
  {Civelli}}, \bibinfo {author} {\bibfnamefont {M.}~\bibnamefont {Capone}},
  \bibinfo {author} {\bibfnamefont {S.~S.}\ \bibnamefont {Kancharla}}, \bibinfo
  {author} {\bibfnamefont {O.}~\bibnamefont {Parcollet}}, \ and\ \bibinfo
  {author} {\bibfnamefont {G.}~\bibnamefont {Kotliar}},\ }\href {\doibase
  10.1103/PhysRevLett.95.106402} {\bibfield  {journal} {\bibinfo  {journal}
  {Phys. Rev. Lett.}\ }\textbf {\bibinfo {volume} {95}},\ \bibinfo {pages}
  {106402} (\bibinfo {year} {2005})}\BibitemShut {NoStop}%
\bibitem [{\citenamefont {Stanescu}\ and\ \citenamefont
  {Kotliar}(2006)}]{Stanescu:2006}%
  \BibitemOpen
  \bibfield  {author} {\bibinfo {author} {\bibfnamefont {T.~D.}\ \bibnamefont
  {Stanescu}}\ and\ \bibinfo {author} {\bibfnamefont {G.}~\bibnamefont
  {Kotliar}},\ }\href {\doibase 10.1103/PhysRevB.74.125110} {\bibfield
  {journal} {\bibinfo  {journal} {Phys. Rev. B}\ }\textbf {\bibinfo {volume}
  {74}},\ \bibinfo {pages} {125110} (\bibinfo {year} {2006})}\BibitemShut
  {NoStop}%
\bibitem [{\citenamefont {Kyung}\ \emph
  {et~al.}(2006{\natexlab{a}})\citenamefont {Kyung}, \citenamefont {Kancharla},
  \citenamefont {S\'en\'echal}, \citenamefont {Tremblay}, \citenamefont
  {Civelli},\ and\ \citenamefont {Kotliar}}]{Kyung:2006a}%
  \BibitemOpen
  \bibfield  {author} {\bibinfo {author} {\bibfnamefont {B.}~\bibnamefont
  {Kyung}}, \bibinfo {author} {\bibfnamefont {S.~S.}\ \bibnamefont
  {Kancharla}}, \bibinfo {author} {\bibfnamefont {D.}~\bibnamefont
  {S\'en\'echal}}, \bibinfo {author} {\bibfnamefont {A.-M.~S.}\ \bibnamefont
  {Tremblay}}, \bibinfo {author} {\bibfnamefont {M.}~\bibnamefont {Civelli}}, \
  and\ \bibinfo {author} {\bibfnamefont {G.}~\bibnamefont {Kotliar}},\ }\href
  {\doibase 10.1103/PhysRevB.73.165114} {\bibfield  {journal} {\bibinfo
  {journal} {Phys. Rev. B}\ }\textbf {\bibinfo {volume} {73}},\ \bibinfo
  {pages} {165114} (\bibinfo {year} {2006}{\natexlab{a}})}\BibitemShut
  {NoStop}%
\bibitem [{\citenamefont {Macridin}\ \emph {et~al.}(2006)\citenamefont
  {Macridin}, \citenamefont {Jarrell}, \citenamefont {Maier}, \citenamefont
  {Kent},\ and\ \citenamefont {D'Azevedo}}]{Macridin:2006}%
  \BibitemOpen
  \bibfield  {author} {\bibinfo {author} {\bibfnamefont {A.}~\bibnamefont
  {Macridin}}, \bibinfo {author} {\bibfnamefont {M.}~\bibnamefont {Jarrell}},
  \bibinfo {author} {\bibfnamefont {T.}~\bibnamefont {Maier}}, \bibinfo
  {author} {\bibfnamefont {P.~R.~C.}\ \bibnamefont {Kent}}, \ and\ \bibinfo
  {author} {\bibfnamefont {E.}~\bibnamefont {D'Azevedo}},\ }\href {\doibase
  10.1103/PhysRevLett.97.036401} {\bibfield  {journal} {\bibinfo  {journal}
  {Phys. Rev. Lett.}\ }\textbf {\bibinfo {volume} {97}},\ \bibinfo {pages}
  {036401} (\bibinfo {year} {2006})}\BibitemShut {NoStop}%
\bibitem [{\citenamefont {Haule}\ and\ \citenamefont
  {Kotliar}(2007)}]{Haule:2007b}%
  \BibitemOpen
  \bibfield  {author} {\bibinfo {author} {\bibfnamefont {K.}~\bibnamefont
  {Haule}}\ and\ \bibinfo {author} {\bibfnamefont {G.}~\bibnamefont
  {Kotliar}},\ }\href {\doibase 10.1103/PhysRevB.76.092503} {\bibfield
  {journal} {\bibinfo  {journal} {Phys. Rev. B}\ }\textbf {\bibinfo {volume}
  {76}},\ \bibinfo {pages} {092503} (\bibinfo {year} {2007})}\BibitemShut
  {NoStop}%
\bibitem [{\citenamefont {Zhang}\ and\ \citenamefont
  {Imada}(2007)}]{Zhang:2007}%
  \BibitemOpen
  \bibfield  {author} {\bibinfo {author} {\bibfnamefont {Y.~Z.}\ \bibnamefont
  {Zhang}}\ and\ \bibinfo {author} {\bibfnamefont {M.}~\bibnamefont {Imada}},\
  }\href {\doibase 10.1103/PhysRevB.76.045108} {\bibfield  {journal} {\bibinfo
  {journal} {Phys. Rev. B}\ }\textbf {\bibinfo {volume} {76}},\ \bibinfo
  {pages} {045108} (\bibinfo {year} {2007})}\BibitemShut {NoStop}%
\bibitem [{\citenamefont {Civelli}\ \emph {et~al.}(2008)\citenamefont
  {Civelli}, \citenamefont {Capone}, \citenamefont {Georges}, \citenamefont
  {Haule}, \citenamefont {Parcollet}, \citenamefont {Stanescu},\ and\
  \citenamefont {Kotliar}}]{Civelli:2008}%
  \BibitemOpen
  \bibfield  {author} {\bibinfo {author} {\bibfnamefont {M.}~\bibnamefont
  {Civelli}}, \bibinfo {author} {\bibfnamefont {M.}~\bibnamefont {Capone}},
  \bibinfo {author} {\bibfnamefont {A.}~\bibnamefont {Georges}}, \bibinfo
  {author} {\bibfnamefont {K.}~\bibnamefont {Haule}}, \bibinfo {author}
  {\bibfnamefont {O.}~\bibnamefont {Parcollet}}, \bibinfo {author}
  {\bibfnamefont {T.~D.}\ \bibnamefont {Stanescu}}, \ and\ \bibinfo {author}
  {\bibfnamefont {G.}~\bibnamefont {Kotliar}},\ }\href {\doibase
  10.1103/PhysRevLett.100.046402} {\bibfield  {journal} {\bibinfo  {journal}
  {Phys. Rev. Lett.}\ }\textbf {\bibinfo {volume} {100}},\ \bibinfo {pages}
  {046402} (\bibinfo {year} {2008})}\BibitemShut {NoStop}%
\bibitem [{\citenamefont {Civelli}(2009)}]{Civelli:2009}%
  \BibitemOpen
  \bibfield  {author} {\bibinfo {author} {\bibfnamefont {M.}~\bibnamefont
  {Civelli}},\ }\href {\doibase 10.1103/PhysRevB.79.195113} {\bibfield
  {journal} {\bibinfo  {journal} {Phys. Rev. B}\ }\textbf {\bibinfo {volume}
  {79}},\ \bibinfo {pages} {195113} (\bibinfo {year} {2009})}\BibitemShut
  {NoStop}%
\bibitem [{\citenamefont {Liebsch}\ and\ \citenamefont
  {Tong}(2009)}]{Liebsch:2009}%
  \BibitemOpen
  \bibfield  {author} {\bibinfo {author} {\bibfnamefont {A.}~\bibnamefont
  {Liebsch}}\ and\ \bibinfo {author} {\bibfnamefont {N.-H.}\ \bibnamefont
  {Tong}},\ }\href {\doibase 10.1103/PhysRevB.80.165126} {\bibfield  {journal}
  {\bibinfo  {journal} {Phys. Rev. B}\ }\textbf {\bibinfo {volume} {80}},\
  \bibinfo {pages} {165126} (\bibinfo {year} {2009})}\BibitemShut {NoStop}%
\bibitem [{\citenamefont {Sakai}\ \emph {et~al.}(2009)\citenamefont {Sakai},
  \citenamefont {Motome},\ and\ \citenamefont {Imada}}]{Sakai:2009}%
  \BibitemOpen
  \bibfield  {author} {\bibinfo {author} {\bibfnamefont {S.}~\bibnamefont
  {Sakai}}, \bibinfo {author} {\bibfnamefont {Y.}~\bibnamefont {Motome}}, \
  and\ \bibinfo {author} {\bibfnamefont {M.}~\bibnamefont {Imada}},\ }\href
  {\doibase 10.1103/PhysRevLett.102.056404} {\bibfield  {journal} {\bibinfo
  {journal} {Phys. Rev. Lett.}\ }\textbf {\bibinfo {volume} {102}},\ \bibinfo
  {pages} {056404} (\bibinfo {year} {2009})}\BibitemShut {NoStop}%
\bibitem [{\citenamefont {Werner}\ \emph {et~al.}(2009)\citenamefont {Werner},
  \citenamefont {Gull}, \citenamefont {Parcollet},\ and\ \citenamefont
  {Millis}}]{Werner:2009}%
  \BibitemOpen
  \bibfield  {author} {\bibinfo {author} {\bibfnamefont {P.}~\bibnamefont
  {Werner}}, \bibinfo {author} {\bibfnamefont {E.}~\bibnamefont {Gull}},
  \bibinfo {author} {\bibfnamefont {O.}~\bibnamefont {Parcollet}}, \ and\
  \bibinfo {author} {\bibfnamefont {A.~J.}\ \bibnamefont {Millis}},\ }\href
  {\doibase 10.1103/PhysRevB.80.045120} {\bibfield  {journal} {\bibinfo
  {journal} {Phys. Rev. B}\ }\textbf {\bibinfo {volume} {80}},\ \bibinfo
  {pages} {045120} (\bibinfo {year} {2009})}\BibitemShut {NoStop}%
\bibitem [{\citenamefont {Gull}\ \emph {et~al.}(2010)\citenamefont {Gull},
  \citenamefont {Ferrero}, \citenamefont {Parcollet}, \citenamefont {Georges},\
  and\ \citenamefont {Millis}}]{Gull:2010}%
  \BibitemOpen
  \bibfield  {author} {\bibinfo {author} {\bibfnamefont {E.}~\bibnamefont
  {Gull}}, \bibinfo {author} {\bibfnamefont {M.}~\bibnamefont {Ferrero}},
  \bibinfo {author} {\bibfnamefont {O.}~\bibnamefont {Parcollet}}, \bibinfo
  {author} {\bibfnamefont {A.}~\bibnamefont {Georges}}, \ and\ \bibinfo
  {author} {\bibfnamefont {A.~J.}\ \bibnamefont {Millis}},\ }\href {\doibase
  10.1103/PhysRevB.82.155101} {\bibfield  {journal} {\bibinfo  {journal} {Phys.
  Rev. B}\ }\textbf {\bibinfo {volume} {82}},\ \bibinfo {pages} {155101}
  (\bibinfo {year} {2010})}\BibitemShut {NoStop}%
\bibitem [{\citenamefont {Lin}\ \emph {et~al.}(2010)\citenamefont {Lin},
  \citenamefont {Gull},\ and\ \citenamefont {Millis}}]{Lin:2010}%
  \BibitemOpen
  \bibfield  {author} {\bibinfo {author} {\bibfnamefont {N.}~\bibnamefont
  {Lin}}, \bibinfo {author} {\bibfnamefont {E.}~\bibnamefont {Gull}}, \ and\
  \bibinfo {author} {\bibfnamefont {A.~J.}\ \bibnamefont {Millis}},\ }\href
  {\doibase 10.1103/PhysRevB.82.045104} {\bibfield  {journal} {\bibinfo
  {journal} {Phys. Rev. B}\ }\textbf {\bibinfo {volume} {82}},\ \bibinfo
  {pages} {045104} (\bibinfo {year} {2010})}\BibitemShut {NoStop}%
\bibitem [{\citenamefont {Sordi}\ \emph {et~al.}(2012)\citenamefont {Sordi},
  \citenamefont {S{\'e}mon}, \citenamefont {Haule},\ and\ \citenamefont
  {Tremblay}}]{Sordi:2012}%
  \BibitemOpen
  \bibfield  {author} {\bibinfo {author} {\bibfnamefont {G.}~\bibnamefont
  {Sordi}}, \bibinfo {author} {\bibfnamefont {P.}~\bibnamefont {S{\'e}mon}},
  \bibinfo {author} {\bibfnamefont {K.}~\bibnamefont {Haule}}, \ and\ \bibinfo
  {author} {\bibfnamefont {A.-M.~S.}\ \bibnamefont {Tremblay}},\ }\href
  {\doibase 10.1038/srep00547} {\bibfield  {journal} {\bibinfo  {journal} {Sci.
  Rep.}\ }\textbf {\bibinfo {volume} {2}} (\bibinfo {year} {2012}),\
  10.1038/srep00547}\BibitemShut {NoStop}%
\bibitem [{\citenamefont {Sordi}\ \emph {et~al.}(2013)\citenamefont {Sordi},
  \citenamefont {S\'emon}, \citenamefont {Haule},\ and\ \citenamefont
  {Tremblay}}]{Sordi:2013}%
  \BibitemOpen
  \bibfield  {author} {\bibinfo {author} {\bibfnamefont {G.}~\bibnamefont
  {Sordi}}, \bibinfo {author} {\bibfnamefont {P.}~\bibnamefont {S\'emon}},
  \bibinfo {author} {\bibfnamefont {K.}~\bibnamefont {Haule}}, \ and\ \bibinfo
  {author} {\bibfnamefont {A.-M.~S.}\ \bibnamefont {Tremblay}},\ }\href
  {\doibase 10.1103/PhysRevB.87.041101} {\bibfield  {journal} {\bibinfo
  {journal} {Phys. Rev. B}\ }\textbf {\bibinfo {volume} {87}},\ \bibinfo
  {pages} {041101} (\bibinfo {year} {2013})}\BibitemShut {NoStop}%
\bibitem [{\citenamefont {Gull}\ \emph {et~al.}(2013)\citenamefont {Gull},
  \citenamefont {Parcollet},\ and\ \citenamefont {Millis}}]{Gull:2013a}%
  \BibitemOpen
  \bibfield  {author} {\bibinfo {author} {\bibfnamefont {E.}~\bibnamefont
  {Gull}}, \bibinfo {author} {\bibfnamefont {O.}~\bibnamefont {Parcollet}}, \
  and\ \bibinfo {author} {\bibfnamefont {A.~J.}\ \bibnamefont {Millis}},\
  }\href {\doibase 10.1103/PhysRevLett.110.216405} {\bibfield  {journal}
  {\bibinfo  {journal} {Phys. Rev. Lett.}\ }\textbf {\bibinfo {volume} {110}},\
  \bibinfo {pages} {216405} (\bibinfo {year} {2013})}\BibitemShut {NoStop}%
\bibitem [{\citenamefont {Gull}\ and\ \citenamefont
  {Millis}(2013)}]{Gull:2013b}%
  \BibitemOpen
  \bibfield  {author} {\bibinfo {author} {\bibfnamefont {E.}~\bibnamefont
  {Gull}}\ and\ \bibinfo {author} {\bibfnamefont {A.~J.}\ \bibnamefont
  {Millis}},\ }\href {\doibase 10.1103/PhysRevB.88.075127} {\bibfield
  {journal} {\bibinfo  {journal} {Phys. Rev. B}\ }\textbf {\bibinfo {volume}
  {88}},\ \bibinfo {pages} {075127} (\bibinfo {year} {2013})}\BibitemShut
  {NoStop}%
\bibitem [{\citenamefont {Imada}\ \emph {et~al.}(2013)\citenamefont {Imada},
  \citenamefont {Sakai}, \citenamefont {Yamaji},\ and\ \citenamefont
  {Motome}}]{Imada:2013}%
  \BibitemOpen
  \bibfield  {author} {\bibinfo {author} {\bibfnamefont {M.}~\bibnamefont
  {Imada}}, \bibinfo {author} {\bibfnamefont {S.}~\bibnamefont {Sakai}},
  \bibinfo {author} {\bibfnamefont {Y.}~\bibnamefont {Yamaji}}, \ and\ \bibinfo
  {author} {\bibfnamefont {Y.}~\bibnamefont {Motome}},\ }\href
  {http://stacks.iop.org/1742-6596/449/i=1/a=012005} {\bibfield  {journal}
  {\bibinfo  {journal} {Journal of Physics: Conference Series}\ }\textbf
  {\bibinfo {volume} {449}},\ \bibinfo {pages} {012005} (\bibinfo {year}
  {2013})}\BibitemShut {NoStop}%
\bibitem [{\citenamefont {Biroli}\ and\ \citenamefont
  {Kotliar}(2002)}]{Biroli:2002}%
  \BibitemOpen
  \bibfield  {author} {\bibinfo {author} {\bibfnamefont {G.}~\bibnamefont
  {Biroli}}\ and\ \bibinfo {author} {\bibfnamefont {G.}~\bibnamefont
  {Kotliar}},\ }\href {\doibase 10.1103/PhysRevB.65.155112} {\bibfield
  {journal} {\bibinfo  {journal} {Phys. Rev. B}\ }\textbf {\bibinfo {volume}
  {65}},\ \bibinfo {pages} {155112} (\bibinfo {year} {2002})}\BibitemShut
  {NoStop}%
\bibitem [{\citenamefont {Aryanpour}\ \emph {et~al.}(2005)\citenamefont
  {Aryanpour}, \citenamefont {Maier},\ and\ \citenamefont
  {Jarrell}}]{Aryanpour:2005}%
  \BibitemOpen
  \bibfield  {author} {\bibinfo {author} {\bibfnamefont {K.}~\bibnamefont
  {Aryanpour}}, \bibinfo {author} {\bibfnamefont {T.~A.}\ \bibnamefont
  {Maier}}, \ and\ \bibinfo {author} {\bibfnamefont {M.}~\bibnamefont
  {Jarrell}},\ }\href {\doibase 10.1103/PhysRevB.71.037101} {\bibfield
  {journal} {\bibinfo  {journal} {Phys. Rev. B}\ }\textbf {\bibinfo {volume}
  {71}},\ \bibinfo {pages} {037101} (\bibinfo {year} {2005})}\BibitemShut
  {NoStop}%
\bibitem [{\citenamefont {Biroli}\ and\ \citenamefont
  {Kotliar}(2005)}]{Biroli:2005}%
  \BibitemOpen
  \bibfield  {author} {\bibinfo {author} {\bibfnamefont {G.}~\bibnamefont
  {Biroli}}\ and\ \bibinfo {author} {\bibfnamefont {G.}~\bibnamefont
  {Kotliar}},\ }\href {\doibase 10.1103/PhysRevB.71.037102} {\bibfield
  {journal} {\bibinfo  {journal} {Phys. Rev. B}\ }\textbf {\bibinfo {volume}
  {71}},\ \bibinfo {pages} {037102} (\bibinfo {year} {2005})}\BibitemShut
  {NoStop}%
\bibitem [{\citenamefont {Maier}\ \emph
  {et~al.}(2005{\natexlab{b}})\citenamefont {Maier}, \citenamefont {Jarrell},
  \citenamefont {Schulthess}, \citenamefont {Kent},\ and\ \citenamefont
  {White}}]{Maier:2005b}%
  \BibitemOpen
  \bibfield  {author} {\bibinfo {author} {\bibfnamefont {T.~A.}\ \bibnamefont
  {Maier}}, \bibinfo {author} {\bibfnamefont {M.}~\bibnamefont {Jarrell}},
  \bibinfo {author} {\bibfnamefont {T.~C.}\ \bibnamefont {Schulthess}},
  \bibinfo {author} {\bibfnamefont {P.~R.~C.}\ \bibnamefont {Kent}}, \ and\
  \bibinfo {author} {\bibfnamefont {J.~B.}\ \bibnamefont {White}},\ }\href
  {\doibase 10.1103/PhysRevLett.95.237001} {\bibfield  {journal} {\bibinfo
  {journal} {Phys. Rev. Lett.}\ }\textbf {\bibinfo {volume} {95}},\ \bibinfo
  {pages} {237001} (\bibinfo {year} {2005}{\natexlab{b}})}\BibitemShut
  {NoStop}%
\bibitem [{\citenamefont {Kyung}\ \emph
  {et~al.}(2006{\natexlab{b}})\citenamefont {Kyung}, \citenamefont {Kotliar},\
  and\ \citenamefont {Tremblay}}]{Kyung:2006b}%
  \BibitemOpen
  \bibfield  {author} {\bibinfo {author} {\bibfnamefont {B.}~\bibnamefont
  {Kyung}}, \bibinfo {author} {\bibfnamefont {G.}~\bibnamefont {Kotliar}}, \
  and\ \bibinfo {author} {\bibfnamefont {A.-M.~S.}\ \bibnamefont {Tremblay}},\
  }\href {\doibase 10.1103/PhysRevB.73.205106} {\bibfield  {journal} {\bibinfo
  {journal} {Phys. Rev. B}\ }\textbf {\bibinfo {volume} {73}},\ \bibinfo
  {pages} {205106} (\bibinfo {year} {2006}{\natexlab{b}})}\BibitemShut
  {NoStop}%
\bibitem [{\citenamefont {Sakai}\ \emph {et~al.}(2012)\citenamefont {Sakai},
  \citenamefont {Sangiovanni}, \citenamefont {Civelli}, \citenamefont {Motome},
  \citenamefont {Held},\ and\ \citenamefont {Imada}}]{Sakai:2012}%
  \BibitemOpen
  \bibfield  {author} {\bibinfo {author} {\bibfnamefont {S.}~\bibnamefont
  {Sakai}}, \bibinfo {author} {\bibfnamefont {G.}~\bibnamefont {Sangiovanni}},
  \bibinfo {author} {\bibfnamefont {M.}~\bibnamefont {Civelli}}, \bibinfo
  {author} {\bibfnamefont {Y.}~\bibnamefont {Motome}}, \bibinfo {author}
  {\bibfnamefont {K.}~\bibnamefont {Held}}, \ and\ \bibinfo {author}
  {\bibfnamefont {M.}~\bibnamefont {Imada}},\ }\href {\doibase
  10.1103/PhysRevB.85.035102} {\bibfield  {journal} {\bibinfo  {journal} {Phys.
  Rev. B}\ }\textbf {\bibinfo {volume} {85}},\ \bibinfo {pages} {035102}
  (\bibinfo {year} {2012})}\BibitemShut {NoStop}%
\bibitem [{\citenamefont {Yin}\ \emph {et~al.}(2014)\citenamefont {Yin},
  \citenamefont {Haule},\ and\ \citenamefont {Kotliar}}]{Yin:2014}%
  \BibitemOpen
  \bibfield  {author} {\bibinfo {author} {\bibfnamefont {Z.~P.}\ \bibnamefont
  {Yin}}, \bibinfo {author} {\bibfnamefont {K.}~\bibnamefont {Haule}}, \ and\
  \bibinfo {author} {\bibfnamefont {G.}~\bibnamefont {Kotliar}},\ }\href
  {http://dx.doi.org/10.1038/nphys3116} {\bibfield  {journal} {\bibinfo
  {journal} {Nat Phys}\ }\textbf {\bibinfo {volume} {10}},\ \bibinfo {pages}
  {845} (\bibinfo {year} {2014})}\BibitemShut {NoStop}%
\bibitem [{\citenamefont {Guterding}\ \emph {et~al.}(2015)\citenamefont
  {Guterding}, \citenamefont {Backes}, \citenamefont {Jeschke},\ and\
  \citenamefont {Valent\'{i}}}]{Valenti:2015}%
  \BibitemOpen
  \bibfield  {author} {\bibinfo {author} {\bibfnamefont {D.}~\bibnamefont
  {Guterding}}, \bibinfo {author} {\bibfnamefont {S.}~\bibnamefont {Backes}},
  \bibinfo {author} {\bibfnamefont {H.~O.}\ \bibnamefont {Jeschke}}, \ and\
  \bibinfo {author} {\bibfnamefont {R.}~\bibnamefont {Valent\'{i}}},\ }\href
  {\doibase 10.1103/PhysRevB.91.140503} {\bibfield  {journal} {\bibinfo
  {journal} {Phys. Rev. B}\ }\textbf {\bibinfo {volume} {91}},\ \bibinfo
  {pages} {140503} (\bibinfo {year} {2015})}\BibitemShut {NoStop}%
\bibitem [{\citenamefont {Aichhorn}\ \emph {et~al.}(2010)\citenamefont
  {Aichhorn}, \citenamefont {Biermann}, \citenamefont {Miyake}, \citenamefont
  {Georges},\ and\ \citenamefont {Imada}}]{Aichhorn:2010}%
  \BibitemOpen
  \bibfield  {author} {\bibinfo {author} {\bibfnamefont {M.}~\bibnamefont
  {Aichhorn}}, \bibinfo {author} {\bibfnamefont {S.}~\bibnamefont {Biermann}},
  \bibinfo {author} {\bibfnamefont {T.}~\bibnamefont {Miyake}}, \bibinfo
  {author} {\bibfnamefont {A.}~\bibnamefont {Georges}}, \ and\ \bibinfo
  {author} {\bibfnamefont {M.}~\bibnamefont {Imada}},\ }\href {\doibase
  10.1103/PhysRevB.82.064504} {\bibfield  {journal} {\bibinfo  {journal} {Phys.
  Rev. B}\ }\textbf {\bibinfo {volume} {82}},\ \bibinfo {pages} {064504}
  (\bibinfo {year} {2010})}\BibitemShut {NoStop}%
\bibitem [{\citenamefont {Tomczak}\ \emph {et~al.}(2012)\citenamefont
  {Tomczak}, \citenamefont {van Schilfgaarde},\ and\ \citenamefont
  {Kotliar}}]{Tomczak:2012}%
  \BibitemOpen
  \bibfield  {author} {\bibinfo {author} {\bibfnamefont {J.~M.}\ \bibnamefont
  {Tomczak}}, \bibinfo {author} {\bibfnamefont {M.}~\bibnamefont {van
  Schilfgaarde}}, \ and\ \bibinfo {author} {\bibfnamefont {G.}~\bibnamefont
  {Kotliar}},\ }\href {\doibase 10.1103/PhysRevLett.109.237010} {\bibfield
  {journal} {\bibinfo  {journal} {Phys. Rev. Lett.}\ }\textbf {\bibinfo
  {volume} {109}},\ \bibinfo {pages} {237010} (\bibinfo {year}
  {2012})}\BibitemShut {NoStop}%
\bibitem [{\citenamefont {Ferrero}\ \emph {et~al.}(2009)\citenamefont
  {Ferrero}, \citenamefont {Cornaglia}, \citenamefont {De~Leo}, \citenamefont
  {Parcollet}, \citenamefont {Kotliar},\ and\ \citenamefont
  {Georges}}]{Ferrero:2009}%
  \BibitemOpen
  \bibfield  {author} {\bibinfo {author} {\bibfnamefont {M.}~\bibnamefont
  {Ferrero}}, \bibinfo {author} {\bibfnamefont {P.~S.}\ \bibnamefont
  {Cornaglia}}, \bibinfo {author} {\bibfnamefont {L.}~\bibnamefont {De~Leo}},
  \bibinfo {author} {\bibfnamefont {O.}~\bibnamefont {Parcollet}}, \bibinfo
  {author} {\bibfnamefont {G.}~\bibnamefont {Kotliar}}, \ and\ \bibinfo
  {author} {\bibfnamefont {A.}~\bibnamefont {Georges}},\ }\href {\doibase
  10.1103/PhysRevB.80.064501} {\bibfield  {journal} {\bibinfo  {journal} {Phys.
  Rev. B}\ }\textbf {\bibinfo {volume} {80}},\ \bibinfo {pages} {064501}
  (\bibinfo {year} {2009})}\BibitemShut {NoStop}%
\bibitem [{\citenamefont {Calder\'on}\ \emph {et~al.}(2009)\citenamefont
  {Calder\'on}, \citenamefont {Valenzuela},\ and\ \citenamefont
  {Bascones}}]{Calderon:2009}%
  \BibitemOpen
  \bibfield  {author} {\bibinfo {author} {\bibfnamefont {M.~J.}\ \bibnamefont
  {Calder\'on}}, \bibinfo {author} {\bibfnamefont {B.}~\bibnamefont
  {Valenzuela}}, \ and\ \bibinfo {author} {\bibfnamefont {E.}~\bibnamefont
  {Bascones}},\ }\href {\doibase 10.1103/PhysRevB.80.094531} {\bibfield
  {journal} {\bibinfo  {journal} {Phys. Rev. B}\ }\textbf {\bibinfo {volume}
  {80}},\ \bibinfo {pages} {094531} (\bibinfo {year} {2009})}\BibitemShut
  {NoStop}%
\bibitem [{Note1()}]{Note1}%
  \BibitemOpen
  \bibinfo {note} {This corresponds to choose $(pd\sigma )^2/|\epsilon _d -
  \epsilon _p|=0.75eV$ in Ref.~\protect \rev@citealp
  {Calderon:2009}.}\BibitemShut {Stop}%
\bibitem [{\citenamefont {Gull}\ \emph {et~al.}(2011)\citenamefont {Gull},
  \citenamefont {Millis}, \citenamefont {Lichtenstein}, \citenamefont
  {Rubtsov}, \citenamefont {Troyer},\ and\ \citenamefont
  {Werner}}]{GullRev:2011}%
  \BibitemOpen
  \bibfield  {author} {\bibinfo {author} {\bibfnamefont {E.}~\bibnamefont
  {Gull}}, \bibinfo {author} {\bibfnamefont {A.~J.}\ \bibnamefont {Millis}},
  \bibinfo {author} {\bibfnamefont {A.~I.}\ \bibnamefont {Lichtenstein}},
  \bibinfo {author} {\bibfnamefont {A.~N.}\ \bibnamefont {Rubtsov}}, \bibinfo
  {author} {\bibfnamefont {M.}~\bibnamefont {Troyer}}, \ and\ \bibinfo {author}
  {\bibfnamefont {P.}~\bibnamefont {Werner}},\ }\href {\doibase
  10.1103/RevModPhys.83.349} {\bibfield  {journal} {\bibinfo  {journal} {Rev.
  Mod. Phys.}\ }\textbf {\bibinfo {volume} {83}},\ \bibinfo {pages} {349}
  (\bibinfo {year} {2011})}\BibitemShut {NoStop}%
\bibitem [{\citenamefont {Haule}(2007)}]{Haule:2007a}%
  \BibitemOpen
  \bibfield  {author} {\bibinfo {author} {\bibfnamefont {K.}~\bibnamefont
  {Haule}},\ }\href {\doibase 10.1103/PhysRevB.75.155113} {\bibfield  {journal}
  {\bibinfo  {journal} {Phys. Rev. B}\ }\textbf {\bibinfo {volume} {75}},\
  \bibinfo {pages} {155113} (\bibinfo {year} {2007})}\BibitemShut {NoStop}%
\bibitem [{\citenamefont {S\'emon}\ \emph {et~al.}(2014)\citenamefont
  {S\'emon}, \citenamefont {Yee}, \citenamefont {Haule},\ and\ \citenamefont
  {Tremblay}}]{Semon:2014}%
  \BibitemOpen
  \bibfield  {author} {\bibinfo {author} {\bibfnamefont {P.}~\bibnamefont
  {S\'emon}}, \bibinfo {author} {\bibfnamefont {C.-H.}\ \bibnamefont {Yee}},
  \bibinfo {author} {\bibfnamefont {K.}~\bibnamefont {Haule}}, \ and\ \bibinfo
  {author} {\bibfnamefont {A.-M.~S.}\ \bibnamefont {Tremblay}},\ }\href
  {\doibase 10.1103/PhysRevB.90.075149} {\bibfield  {journal} {\bibinfo
  {journal} {Phys. Rev. B}\ }\textbf {\bibinfo {volume} {90}},\ \bibinfo
  {pages} {075149} (\bibinfo {year} {2014})}\BibitemShut {NoStop}%
\bibitem [{\citenamefont {Luttinger}\ and\ \citenamefont
  {Ward}(1960)}]{LuttingerWard:1960}%
  \BibitemOpen
  \bibfield  {author} {\bibinfo {author} {\bibfnamefont {J.~M.}\ \bibnamefont
  {Luttinger}}\ and\ \bibinfo {author} {\bibfnamefont {J.~C.}\ \bibnamefont
  {Ward}},\ }\href {\doibase 10.1103/PhysRev.118.1417} {\bibfield  {journal}
  {\bibinfo  {journal} {Phys. Rev.}\ }\textbf {\bibinfo {volume} {118}},\
  \bibinfo {pages} {1417} (\bibinfo {year} {1960})}\BibitemShut {NoStop}%
\bibitem [{\citenamefont {Nomura}\ \emph {et~al.}(2015)\citenamefont {Nomura},
  \citenamefont {Sakai},\ and\ \citenamefont {Arita}}]{Nomura:2015}%
  \BibitemOpen
  \bibfield  {author} {\bibinfo {author} {\bibfnamefont {Y.}~\bibnamefont
  {Nomura}}, \bibinfo {author} {\bibfnamefont {S.}~\bibnamefont {Sakai}}, \
  and\ \bibinfo {author} {\bibfnamefont {R.}~\bibnamefont {Arita}},\ }\href
  {\doibase 10.1103/PhysRevB.91.235107} {\bibfield  {journal} {\bibinfo
  {journal} {Phys. Rev. B}\ }\textbf {\bibinfo {volume} {91}},\ \bibinfo
  {pages} {235107} (\bibinfo {year} {2015})}\BibitemShut {NoStop}%
\end{thebibliography}

\begin{thebibliography}{53}%
\makeatletter
\providecommand \@ifxundefined [1]{%
 \@ifx{#1\undefined}
}%
\providecommand \@ifnum [1]{%
 \ifnum #1\expandafter \@firstoftwo
 \else \expandafter \@secondoftwo
 \fi
}%
\providecommand \@ifx [1]{%
 \ifx #1\expandafter \@firstoftwo
 \else \expandafter \@secondoftwo
 \fi
}%
\providecommand \natexlab [1]{#1}%
\providecommand \enquote  [1]{``#1''}%
\providecommand \bibnamefont  [1]{#1}%
\providecommand \bibfnamefont [1]{#1}%
\providecommand \citenamefont [1]{#1}%
\providecommand \href@noop [0]{\@secondoftwo}%
\providecommand \href [0]{\begingroup \@sanitize@url \@href}%
\providecommand \@href[1]{\@@startlink{#1}\@@href}%
\providecommand \@@href[1]{\endgroup#1\@@endlink}%
\providecommand \@sanitize@url [0]{\catcode `\\12\catcode `\$12\catcode
  `\&12\catcode `\#12\catcode `\^12\catcode `\_12\catcode `\%12\relax}%
\providecommand \@@startlink[1]{}%
\providecommand \@@endlink[0]{}%
\providecommand \url  [0]{\begingroup\@sanitize@url \@url }%
\providecommand \@url [1]{\endgroup\@href {#1}{\urlprefix }}%
\providecommand \urlprefix  [0]{URL }%
\providecommand \Eprint [0]{\href }%
\providecommand \doibase [0]{http://dx.doi.org/}%
\providecommand \selectlanguage [0]{\@gobble}%
\providecommand \bibinfo  [0]{\@secondoftwo}%
\providecommand \bibfield  [0]{\@secondoftwo}%
\providecommand \translation [1]{[#1]}%
\providecommand \BibitemOpen [0]{}%
\providecommand \bibitemStop [0]{}%
\providecommand \bibitemNoStop [0]{.\EOS\space}%
\providecommand \EOS [0]{\spacefactor3000\relax}%
\providecommand \BibitemShut  [1]{\csname bibitem#1\endcsname}%
\let\auto@bib@innerbib\@empty
\bibitem [{\citenamefont {Biroli}\ \emph {et~al.}(2004)\citenamefont {Biroli},
  \citenamefont {Parcollet},\ and\ \citenamefont {Kotliar}}]{Biroli:2004}%
  \BibitemOpen
  \bibfield  {author} {\bibinfo {author} {\bibfnamefont {G.}~\bibnamefont
  {Biroli}}, \bibinfo {author} {\bibfnamefont {O.}~\bibnamefont {Parcollet}}, \
  and\ \bibinfo {author} {\bibfnamefont {G.}~\bibnamefont {Kotliar}},\ }\href
  {\doibase 10.1103/PhysRevB.69.205108} {\bibfield  {journal} {\bibinfo
  {journal} {Phys. Rev. B}\ }\textbf {\bibinfo {volume} {69}},\ \bibinfo
  {pages} {205108} (\bibinfo {year} {2004})}\BibitemShut {NoStop}%
\end{thebibliography}
\end{document}